\documentclass[12pt]{article}
\usepackage{amssymb,latexsym}
\usepackage{graphicx}
\usepackage{amssymb, amsmath, amsopn, amsthm,slashed}
\usepackage{epsfig}

\def\be{\begin{equation}}
\def\ee{\end{equation}}
\def\bea{\begin{eqnarray}}
\def\eea{\end{eqnarray}}

\def\K {K\"{a}hler}

\begin {document}
\begin{titlepage}
\hfill SU-ITP-07/16\\

\begin{center}

\vskip 2cm

{ \Large \bf     Axion Inflation and \\

\

Gravity Waves in String Theory}

\vskip 1cm

{\bf
Renata Kallosh\footnote{\mbox{
Email: {\tt kallosh@stanford.edu} }},
Navin
Sivanandam\footnote{\mbox{
Email: {\tt navins@stanford.edu} }}
and Masoud Soroush\footnote{\mbox{
Email: {\tt soroush@stanford.edu}
}}}
\\

\vskip 1cm


{\em }

\centerline{\textsl{Institute for Theoretical Physics, Department of Physics}}
\centerline{\textsl{Stanford University, Stanford, CA 94305-4060, USA}}
\centerline{}

 {\bf ABSTRACT }
 \end{center}
The  majority of models of inflation in string theory predict an absence of measurable  gravitational waves, $r << 10^{-3}$. The most promising proposals for making string theoretic models that yield measurable tensor fluctuations involve axion fields with slightly broken shift symmetry. We consider such models in detail, with a particular focus on the N-flation scenario and on axion valley/natural inflation models. We find that in Calabi-Yau threefold compactifications with logarithmic K\"{a}hler potentials $K$ it appears to be difficult to meet the conditions required for axion inflation in the supergravity regime. However, in supergravities with an (approximately) quadratic shift-symmetric $K$, axion inflation may be viable. Such K\"ahler potentials do arise in some string models, in specific limits of the moduli space. We describe the most promising classes of models; more detailed study will be required before one can conclude that working models exist.

\vfill

\end{titlepage}
\tableofcontents{}
\section{Introduction}
The space of inflationary models is vast and varied. While it may be impossible to reconstruct the precise inflationary potential from data, narrowing the classes of models to those that give rise to our corner of the universe is a vital step in describing our cosmology. A promising avenue for doing so in the near future will be the measurement of the level of tensor fluctuations from inflation \cite{Efstathiou:2006ak}, which is typically
characterized by the ratio $r=T/S$, where $T$ and $S$ are the amplitudes of tensor and scalar perturbations respectively

Experimental constraints on $r$ come from measurements of the microwave background polarization.  The current bound from WMAP and SDSS is $r < 0.3$ \cite{Tegmark:2006az}. We can expect this limit on $r$ to be driven down by the next generation of CMB experiments. In addition to the soon to be launched Planck satellite, there are quite a few experiments on the horizon that have been specifically designed to measure B-mode polarization and thus inflationary gravity waves (IGW).\footnote{In this paper  we consider only B-modes from tensor fluctuations during inflation. We will not discuss B-modes from cosmic strings,  which might be produced during the exit from inflation. The observational signatures for these two types of B-modes are quite different, see for example \cite{Pogosian:2006hg}.} These include Spider, SPUD, EBEX, polarBear, QUIET, and Clover. BICEP and QUAD are already taking data now. All of these hope to detect $r \geq 0.1$ (or even $r \geq 0.02$ for BICEP2) or place the corresponding upper bound, and should have results by 2011 \cite{Netterfield}. Further into the future the NASA CMBpol satellite project \cite{Bock:2006yf} is expected to be launched by 2018 and hopes to achieve a detection or a bound at $r < ~0.01$. The ESA Bpol project has an analogous goal.

There are known inflationary models predicting detectable levels of IGW with $ 10^{-3} <r < 0.3$. These include chaotic inflationary models \cite{Linde:1983gd,Destri:2007pv,Kallosh:2007wm,Linde:2007fr}, with quadratic, cubic and quartic potentials where one expects $r\geq 2\cdot 10^{-2}$ \cite{Destri:2007pv}, and natural inflation models \cite{Freese:1990rb} where one expects $r\geq 10^{-3}$, \cite{Savage:2006tr}.  The potential for natural inflation is $\Lambda^4(1-\cos (\phi/f))$ where $\phi$ is the canonically normalized field and $f$ is the axion decay constant.  Quite generally, as shown by Lyth (and exemplified by chaotic inflation), single-field models predicting measurable IGW require a super-Planckian variation of the inflaton over the 60 e-foldings corresponding to the observable part of our universe; $\Delta \phi >  M_{Pl}$ \cite{Lyth:1996im}. For axion models this implies $f > M_{Pl}$ \footnote{We use the reduced Planck mass, $M_{Pl}=(\sqrt {8\pi G_N})^{-1}$}.  The variation of individual fields can be somewhat smaller in models of assisted inflation \cite{Liddle:1998jc}, with many fields increasing the frictional force during inflation. In such models the value of the VEVs of individual fields may be reduced; for example by a factor of ${1\over \sqrt N}$ in assisted $m^2\phi^2$ inflation, where $N$ is the number of fields.

Our concern in this paper is the study of models derivable from string theory and effective supergravity which have appreciable levels of IGW. Such models are rare and, as discussed recently in \cite{Kallosh:2007wm,Kachru:unpub,Kallosh:2007ig}, they are not amongst those scenarios that have been well studied in string theory. For example, brane inflation models were shown to be IGW free in \cite{Baumann:2006cd,Bean:2007hc} and the same is true of modular inflation proposals (see \cite{Kallosh:2007ig} for a review of these models).

Part of the problem with embedding IGW in string theory is that the value of $r$ correlates directly with the energy scale of inflation. For a given value of $r$, the Hubble constant during inflation is roughly $H\sim r^{1/2}\times 10^{14} ~{\rm GeV}$. This dependence means that, for measurable values of $r$, inflation would have to have occurred at close to GUT-scale energy densities -- $V \sim (10^{16} {\rm GeV})^4$. Now, since the moduli of a string theory compactification need to be stabilized above the energy scale of inflation to obtain a robust model (as discussed in many papers such as \cite{Kachru:2003sx}), it is clear that generating a measurable value $r$ translates into the need for moduli stabilization at very high energy scales.

Unfortunately, achieving this is a challenging problem. In the simplest methods of moduli stabilization, such as the one advocated in the KKLT
paper \cite{Kachru:2003aw},
 there is a constraint relating the mass of
gravitino to the Hubble parameter, $m_{3/2}\geq H$
\cite{Kallosh:2004yh}. This would predict $r\sim 10^{-24}$ for
$m_{3/2} \sim  \textrm{TeV}$ and require an extremely heavy
gravitino for detection at $r\sim 10^{-2}$, as shown in
\cite{Kallosh:2007wm}. It is possible to avoid this problem by modifying the original KKLT construction as shown in \cite{Kallosh:2004yh}. However, achieving $r \sim 10^{-2}$ in  such models for
$m_{3/2} \sim  \textrm{TeV}$ would require significant fine-tuning \cite{Kallosh:2007wm}.

An additional difficulty with finding IGW in string theory arises from considering the aforementioned Lyth bound (the general requirement of trans-Planckian field motion to get measurable gravity waves) in string theory. Not much is known about the behavior of a variation of a scalar modulus over a range $\Delta \phi > M_{\rm string}$ in a string theory moduli space and what we suspect is not encouraging. In general, maintaining the flatness of the potential over a distance $\Delta \phi \gg M_{\rm string}$  appears to be difficult. However, by using axions, whose potential is to a significant extent determined by symmetries, one would evade some of the issues that appear to prevent flatness \cite{Binetruy:1986ss,Gaillard:1995az}. Unfortunately, though, it was pointed out in \cite{Banks:2003sx} that the super-Planckian axion decay constants $f > M_{Pl}$ required for desired regime seem to be unavailable in controlled limits of string theory. While this constraint is met by many of the models we consider in this paper, we note that the argument (reviewed in section \ref{Banks}) has been shown in \cite{Kim:2004rp} to be avoidable for a particular choice of two axions. Furthermore, we will demonstrate in section \ref{sugramulti} that it is possible to avoid the problem even for one axion field, with a particular racetrack-type superpotential. In addition, we will discuss some  numerical factors of $2^m$ and $\pi^n$ which need clarification to make theoretical string cosmology compatible with cosmological precision data.

Even without $f > M_{Pl}$ axions can still provide IGW-generating potential through the assistance effect discussed above. Assisted inflation has been implemented with stringy axions in the recent N-flation scenario \cite{Dimopoulos:2005ac}. As mentioned previously, in assisted models one tries to benefit from the Hubble friction generated by N different fields rolling towards their minima. If one can enforce, via symmetries, that the cross-couplings between the fields are suppressed, then one finds that the inflationary slow roll parameters scale with inverse powers of $N$. It was argued in \cite{Dimopoulos:2005ac} that string axions provide excellent candidates for assisted inflation.  In many models, they are numerous (with the number being determined by the topology of the compact manifold) and they have independent shift symmetries which are broken by distinct instantons.  This provides a rationale for the lack of cross-couplings.  For sufficiently large N, and with a proper choice of scales, one can then hope to make a stringy model of axion assisted inflation that generates IGW. However, although N-flation circumvents several obstacles to inflationary model-building in string theory, it does so at the cost of making certain assumptions. One of our goals in this paper is to see if these assumptions can be realized in controlled string compactifications.

One of the main assumptions in the analysis of \cite{Dimopoulos:2005ac} was that the scalar superpartners of the axions (which are K\"ahler moduli in many classes of models), get effective masses $>H$ during inflation and decouple from the dynamics. Accordingly, we will concern ourselves with the necessity of finding models of moduli stabilization where the volume moduli have much steeper potentials than their axionic partners. If such a model can be found, then quite plausibly the physics will be that of many massive axions acting in concert to give some combination of assisted chaotic inflation and natural inflation. If not, while there is no general proof that inflation does not occur, the story is certainly more complicated than the one described in \cite{Dimopoulos:2005ac}.

In \cite{Kallosh:2007ig} a class of ${\cal N}=1$ supergravity potentials with this mass hierarchy (for the specific case of a single complex modulus) were described, and given the name ``axion valley'' potentials. A natural inflation model \cite{Freese:1990rb} with a pseudo-Nambu-Goldstone boson (pNGb) is a slice of the bottom of the 2d axion valley potential of the supergravity model and, very roughly speaking, an N-flation model involves decoupled dynamics along the bottom of $N$ such valleys. Thus if axion valleys exist in the string landscape we can obtain measurable IGW by realizing either pNGb inflation or N-flation -- the latter has the obvious advantage of having only sub-Planckian VEVs. In \cite{Kallosh:2007ig} the preliminary analysis suggested that such valleys are hard to find for axions that are superpartners of K\"ahler moduli in Calabi-Yau threefolds (or orientifolds),
but they may exist for axions or brane position modes which enjoy quadratic shift symmetric K\"{a}hler potentials (in some approximation scheme). Here we will make a thorough analysis of this issue, describing some classes of known models which can give rise to the promising K\"ahler potentials and test carefully several possibilities for deriving appropriate axion valley potentials from string theory.

Our explorations of axions and inflation in string theory will follow a structure based on the observations made above. We begin in section  \ref{nflation} by considering the status of axions in string theory and the apparent restrictions on their decay constants. We also discuss the details of N-flation in more depth. We then loosely divide the models we consider into the following categories. First, in section \ref{onemodulus}, we consider models with K\"{a}hler moduli that have a shift symmetric logarithmic K\"{a}hler potential (associated with the cubic prepotential) and a non-perturbative superpotential, such as one would expect in KKLT-type scenarios in IIB theories. We demonstrate, in a number of examples, that both the  axion and its partner get masses of the same order -- i.e. we fail to meet the criteria for N-flation;\, $N$ moduli would not give $N$ decoupled axions.

In section \ref{kahlerstab} we consider similar models, but use a more general form for the K\"{a}hler potential, with $\alpha'$ corrections to the tree level form. Naively this K\"{a}hler potential is such that one might expect it stabilize the volume part of the moduli, leaving a relatively flat axion direction for each modulus, even with a constant superpotential. However, by examining the scenario in detail we prove that stabilization cannot take place except far in the interior of moduli space, where our formulae are suspect and we have little control. We follow this analysis in section \ref{othermodels}, by looking at related examples in the literature, but with a non-trivial superpotential. Here, though the situation is somewhat different, the examples we study do not provide us with the framework necessary for assisted axion inflation.

We continue our studies in section \ref{sugramodels} where we first describe the supergravity axion valley model proposed in \cite{Kallosh:2007ig},  which leads to a natural axion inflation \cite{Freese:1990rb} due to a quick stabilization of a radial partner of the axion-inflaton. We present an example of an ``racetrack fix'' mechanism  for this model where by using two exponents in the superpotential one can avoid an argument of  \cite{Banks:2003sx} which puts a constraint on the axion decay constant. We further look at some models in string theory which may lead to a regime of the effective supergravity of the type required in \cite{Kallosh:2007ig} and conclude that more detailed studies will be required.

In section \ref{discussion} we stress that investigation  of IGW in string cosmology is very complicated. We have just began the exploration of string theory models where one is able to perform an analysis. Although the best studied models fail to lead to IGW, we do have some indications of classes of models where this is not the case. These models are less understood and require further study, but may well provide predictions of a measurable scalar-tensor ratio.

\section{Axions and Assisted Inflation in String Theory}\label{nflation}
\subsection{Super-Planckian decay constant for a single axion}\label{Banks}
In the simplest context, that of a non-supersymmetric theory with an axion field $\phi$ of decay constant $f$ and Peccei-Quinn breaking scale $\Lambda$, the potential takes the form
\begin{equation}
\label{axpot}
V = \Lambda^4 (1 - \cos(\phi/f))~.
\end{equation}
Assuming canonical kinetic terms ${1\over 2} (\partial \phi)^2$, it follows that the  tensor to scalar ratio $r$ defined by the slow-roll parameter $\epsilon$ is
\begin{equation}
\label{scaling}
r= 16 \epsilon = 8 (M_{Pl}^2 \left({V'\over V}\right)^2= 8 \left({M_{Pl}\over f} \right)^2 \left({\sin(\phi/f))\over (1 - \cos(\phi/f))}\right)^2 \ .
\end{equation}
The actual value of the field $\phi$ where IGW are produced depends on $f$. Numerical computations in \cite{Savage:2006tr} show that natural inflation with $r>5\times 10^{-3}$ is possible for $f> 4M_{Pl}$.

Various ideas for deriving four-dimensional effective theories with axions having super-Planckian $f$ have been proposed over the years; one recent interesting set of ideas appeared in \cite{ArkaniHamed:2003wu}. However, it is thought that string theory does not allow (parametrically) super-Planckian $f$s, as argued in \cite{Banks:2003sx}. This conclusion also follows from the concrete formulae derived for axion decay constants in the known weak-coupling limits of string theory in \cite{Svrcek:2006yi}. Here, we briefly review the argument of \cite{Banks:2003sx}, to justify our partial focus on the admittedly more baroque multifield models in the bulk of this paper.

One starts with a modular invariant action of the type
\begin{equation}
L= c {(\partial \sigma)^2 + (\partial \beta)^2\over 4 \sigma^2} -
V(\sigma, \beta) \ .
\end{equation}
Here $\sigma$ is one of the moduli, for example the volume or the dilaton, and $\beta$ is the axion partner of the corresponding volume, $\rho= \sigma +i \beta$. The K\"{a}hler potential is of the form $K= -c\ln (\rho +\bar \rho)$. In type IIB Calabi-Yau compactifications, for the dilaton $c=1$, while for the total volume $c=3$. It is assumed that the dilaton/volume modulus is stabilized at some $\sigma=\sigma_{0}$, and that the potential depends
on the periodic axion due to instanton corrections
\begin{equation}
V\sim \cos \beta ~.
\end{equation}

The kinetic term for the axion at the fixed values of $\sigma$ is given by
\begin{equation}
{1\over 4} {c\over \sigma_0^2} (\partial\beta)^2 = {1\over 2}
(\partial \tilde \beta)^2
\end{equation}
Here $\tilde \beta$ is the axion with a canonical kinetic term. The potential depends on $\cos\beta=\cos (\sigma_0\tilde\beta/\sqrt{2c})= \cos (\tilde\beta/f)$. The axion decay constant is therefore inversely proportional to the stabilized value of the dilaton/volume:
\begin{equation}
f= {\sqrt{c/2}\over \sigma_0} ~.\label{f}
\end{equation}

Meanwhile, the axion potential arises from Euclidean branes wrapped over some cycles (whose volume is controlled by $\sigma_0$):
\begin{equation}
e^{-(\sigma +i\beta)}|_{\sigma_0}\sim e^{-(\sigma_0 +i {\sigma_0
\tilde \beta \over\sqrt{c/2}})} \label{inst}\sim e^{-(\sigma_0 +i {
\tilde \beta \over f})}
\end{equation}
It follows from (\ref{f}) that large decay constant implies small $\sigma_0$. Then, according to (\ref{inst}), the term $e^{-\sigma_0}$ is
not very small and therefore the n-instanton corrections proportional to $e^{-n \sigma_0}$ must also be taken into account.  More detailed
analysis reveals that, for all $f$, all harmonics up to $n \sim (f/M_{Pl})$ give an appreciable contribution. This presents a serious obstruction to the possibility of having parametrically large axion decay constants in string theory, since the multi-instanton corrections will destroy the desired large periodicity of the single instanton contribution, reducing the ``effective" $f$ by a factor of $n$.

\subsubsection{A racetrack fix}
In the considerations above it was implicitly assumed that there is only one exponential term in the superpotential, originating from an instanton or gaugino condensate superpotential, of the form
\begin{equation}
W\sim  e^{-\rho}
\end{equation}
Meanwhile, one can look at the racetrack models with two exponents (due, for example, to two different gauge groups) and obtain:
\begin{equation}
W=  A e^{-2\pi\rho\over N}+ B e^{-2\pi\rho\over M} \ .
\end{equation}
In such a case the potential depends on \cite{BlancoPillado:2004ns}
\begin{equation}
V\sim \cos{(N-M) \beta\over MN}
\end{equation}
and the relevant axion decay coupling is
\begin{equation}
f_{racetrack} \approx {MN \over (M-N)\sigma_0} ~.\label{f1}
\end{equation}
At large $M$ and $N$ and small $M-N$ it may be possible to get large $f$ without a small $\sigma_0$. In this way the racetrack type models can eliminate the problem discussed in \cite{Banks:2003sx}. A related solution for the case of 2 axions was proposed and studied  in detail in \cite{Kim:2004rp}.

\subsection{The issue of  $2^m \pi^n$ factors in string theory}\label{twopi}
In the paper \cite{Banks:2003sx} there is a remark that some estimates concerning the status of the axion decay constant in string theory may be modified by relatively large factors like $16\pi^2$. We find this observation particularly important for  generic situations in string cosmology. We came across analogous problems in previous attempts to relate string theory with observational cosmology.

At the time of precision cosmology the clarification of the factor $\sqrt{8\pi}$  between the old Planck mass, $M^{old}_{Pl}= {1\over  G_N}= 1.221 \times 10^{19}$GeV and ``new'' or ``reduced'' Planck mass $M_{Pl}= {M^{old}_{Pl}\over \sqrt {8\pi}}= 2.436\times 10^{18}$GeV become very important. The origin of this difference is in normalization of the Einstein term in the action of 4d general relativity
\begin{eqnarray}
S_{4d}&= &{1\over 16\pi G_{N}} \int d^4 x \, \sqrt{-g} R = {(M_{Pl}^{old})^2\over 16 \pi}\int d^4 x \, \sqrt{-g} R \nonumber\\
&= &{(M_{Pl})^2\over 2}\int d^4 x \, \sqrt{-g} R = {1\over 2
\kappa^2_4}\int d^4 x \, \sqrt{-g} R \ . \label{4d}
\end{eqnarray}
The reason for calling the reduced Planck mass a ``new'' Planck mass has to do with the fact that in supergravity literature one typically finds actions in the form
\begin{equation}
S_{supergravity}= {1\over 2} \int d^4 x \, \sqrt{-g} R +... \qquad \Rightarrow \qquad M_{Pl}^{new}=1
\end{equation}
The Planck length also has an ``old'' and a ``new'' value
\begin{equation}
l_{Pl}^{old}= 1.616\times 10^{-33} \rm cm \ , \qquad l_{Pl}^{new}=\sqrt
{8\pi} l_{Pl}^{old} 0.81\times 10^{-32} \rm cm
\label{l}\end{equation}

Clearly, a factor like ${8\pi}\sim 25$ cannot be neglected when it enters the definition of observables like a dimensional slow roll parameters in inflation
\begin{equation}
\eta= (M_{Pl})^2 {V''\over V}= {1\over 8\pi}(M_{Pl}^{old})^2
{V''\over V} \ , \qquad \epsilon = {(M_{Pl})^2\over 2} \left(
{V'\over V}\right)^2={1\over 8\pi} {(M_{Pl}^{old})^2\over 2} \left(
{V'\over V}\right)^2
\end{equation}
The current observed value of the spectral index $n_s$ is given by \cite{Tegmark:2006az}
\begin{equation}
n_s= 1+  2\eta -6\epsilon= 0.953\pm 0.016
\end{equation}
Therefore the calculation of the values of $n_s-1$ from the theory should not be off by a factor of $8\pi\sim 25$ (which it would be if the difference between units $M_{Pl}^{new}=1$ and $M_{Pl}^{old}=1$ would not be clarified in all current cosmology papers and books). The important fact about the existence of the difference between old cosmological parameters and reduced ones originating from supergravity applications to cosmology is the following: one can use any of these parameters and work with any choice of units as long as the final numerical theoretical value is related to observations in a way which does not depend on the choice of units. In string theory the analogous has not been clarified, as the above quote from \cite{Banks:2003sx} demonstrates and the recent paper \cite{Svrcek:2006yi} confirms.

What is the relation between $l_{Pl}$ and $l_{st}$? As  explained above, one may use in this relation either old or new $l_{Pl}$, there is a simple relation between them shown in eq. (\ref{l}). However, what precisely is $l_{st}$? We may start with the definition of string length $l_s$ and string
units $l_s=1$  in various papers and books in string literature. In the first part of the book \cite{Becker:2007zj} we find
\begin{equation}
l_s = \sqrt{ 2\alpha'} \qquad T= {1\over 2\pi \alpha'}
\end{equation}
Here $\alpha'$ is  the open string Regge-slope parameter related to the string tension parameter $T$. The relevant world-sheet Nambu-Goto action is
\begin{equation}
S_{2d}= -{T\over 2} \int d\tau d\sigma \sqrt{-h} h^{\alpha \beta}
\eta_{\mu\nu} \partial_\alpha X^\mu \partial_\beta X^\nu
\end{equation}
and the units
\begin{equation}
\alpha'={1\over 2} \qquad l_s=1 \qquad T={1\over \pi}
\end{equation}
are used. In the part of the book which deals not with the world-sheet but with effective eleven or ten-dimensional supergravity a slightly different definition of string length is used
\begin{equation}
\tilde l_s=\sqrt {\alpha'}= {l_s\over  \sqrt{ 2}}
\end{equation}
Thus for the volume of the extra dimensions measured in string units we may easily encounter the factor of 8  difference depending on units which are used:
\begin{equation}
V_6\sim l_s^6 = 2^3 \tilde l_s
\end{equation}

In \cite{Polchinski:1998rr} a detailed presentation of the D-brane physics is given and we would like to point out various sources of difficulties with regard to powers of $2$ and $\pi$ when applied to 4d cosmology. The most relevant formulas are for the value of the Planck mass in the 10d string frame
\begin{equation}
k_{10}^2= {1\over 2} (2\pi)^7 (\alpha')^4 \label{10d}
\end{equation}
where the gravitational part of the 10d action is
\begin{equation}
S_{10d} = {1\over 2k_{10}^2}\int d^{10} x e^{-2\Phi} \sqrt{-G} R \ ,
\end{equation}
and the tension of the $D_p$-branes is
\begin{equation}
\mu_p = {1\over (2\pi)^p (\alpha')^{p+1\over 2}} \ .
\label{tension}
\end{equation}
The $D_p$- brane action including Born-Infeld term and Chern-Simons term is
\begin{equation}
D_p= \mu_p \left [-\int d^{p+1} \xi e^{-\Phi}\sqrt {-\det [\tilde
G_{ab}]}+ \int C_{p+1}\right] \ .
\end{equation}

Here there are few general remarks in order. The equations for the gravitational coupling (\ref{10d}) and the brane tension (\ref{tension}) were derived in \cite{Polchinski:1998rr} strictly for toroidal compactification of 10d theory to 4d one. In particular (ignoring for the moment the issue of the string coupling) a dimensional reduction on a product of six circles of radius $r$ will lead to

\begin{equation}
S_{10d} = {1\over (2\pi)^7 (\alpha')^4}\int d^{10} x e^{-2\Phi}
\sqrt{-G} R_{10} \Rightarrow {(2\pi r)^6\over (2\pi)^7 (\alpha')^4}\int
d^{4}x  \sqrt{-g} R_4
\end{equation}
Comparing this with (\ref{4d}) we find that the 4d Planck length is related to the string length $\sim \sqrt {\alpha'}$ and the radius of compactification $r$ as follows:
\begin{equation}
l_{Pl}^{old}= \sqrt{ \alpha'} \left ({\sqrt
{\alpha'}\over {\sqrt 2} r} \right)^3
\end{equation}
The large factors $(2\pi)^6$ cancel and for ${\sqrt{\alpha'}\over {\sqrt 2} r}\sim1$ the old Planck length and the string length are equal. Meanwhile, relations of the type (\ref{10d}) have been applied to Calabi-Yau compactifications with volumes different from toroidal compactifications. To stress the numerical importance of such differences we may compare volumes for toroidal, $(2\pi r)^6$, and spherical, ${\pi^3\over 6}r^6$ compactifications:
\begin{equation}
{V_{6-circles}\over V_{6-sphere}}= { 6(2\pi)^6\over \pi^3}= 48 (2\pi)^3 \sim 10^4
\end{equation}
Thus one can significantly change the relation between the string length and the 4d Planck length, keeping $r=\sqrt {\alpha'}$, but using different numerical values for the volume of the internal space. Moreover, it has not been established how to change equations (\ref{10d}) and (\ref{tension}) for non-toroidal compactifications.

Another sensitive point of our discussion, apart from the 10d/4d relation is the identification of the scale of non-perturbative corrections to the superpotential. This will be related to the axions in string theory and the $16\pi^2$ or so factors for the axion decay constant. For example, in \cite{Svrcek:2006yi} and\cite{Denef:2004dm} the string length is defined as
\begin{equation}
\hat l_s= 2\pi \sqrt {\alpha'}
\end{equation}
and the brane tension is
\begin{equation}
\mu_p= {2\pi\over (2 \pi \sqrt{\alpha'})^{p+1}}= {2\pi\over (\hat l_s)^{p+1}} \ .
\end{equation}
These conventions lead to a significant simplification since the appearance of the factors $2\pi$ is minimized. In particular, the action of the D-p-brane has a simple Chern-Simons form term in $\hat l_s=1$ units
\begin{equation}
S_{C-S}= 2\pi \int C_4
\end{equation}
and $p$-form fields have integer periods. In these conventions it is easy to see that quantum corrections break $SL(2.\mathbb{R}$ symmetry of the supergravity to $SL(2.\mathbb{Z})$ symmetry of string theory. Indeed, the Euclidean action of the D-3-brane, for example, is given by \cite{Denef:2004dm}
\begin{equation}
e^{-2\pi  \tau} \label{DDF}
\end{equation}
where $ \tau$ is the complex volume-axion modulus
\begin{equation}
\tau= \int_{D} {1\over 2} J\wedge J -i C_4
\end{equation}
This simple expression is valid only in  $\hat l_s=1$ units. In general, one should use
\begin{equation}
e^{-{2\pi \over (2 \pi \sqrt{\alpha'})^{4}}\tau} \label{general}
\end{equation}
Depending on units in which the volume modulus $\tau$ is measured, the instanton action would have a factor of $(2\pi)^4$ difference with eq. (\ref{DDF}) for example if $\alpha'=1$ units were used.

To make things even more complicated we note that the stringy units used in \cite{Becker:2002nn} are $2\pi\alpha'={(\hat l_s)^2\over2\pi}=1$ which is different from all previous cases. With regard to the axion decay constant ambiguity remains in extracting the kinetic term for the modulus and translating to 4d units with a specific relation between string units and $M_{Pl}$.

The instanton corrections to the superpotential depend dramatically on careful consistent numerical computations in string theory models. Note that the difference between the detection of IGW and non-detection with regard to a single exponent axion model proposed in \cite{Kallosh:2007ig} with $W\sim e^{-b \Phi}$ boils down to the difference between $b\sim 0.28$ and $b\sim 0.04$. Clearly the ambiguity in $b$ of the order $16\pi^2$ noticed in
\cite{Banks:2003sx} and explained in detail here is not acceptable if the prediction of the theory is to be compared with the data.

We presented here few situations in string cosmology where the factors of $2^n \pi^m$ may need clarification. In particular, with regard to instanton corrections to the superpotential we will make choices of the coefficient in the exponent in the superpotential which will give the phenomenology of inflation with and without gravity waves. The actual numbers will be of crucial importance.

\subsection{On N-flation: assisted inflation in string theory}
As explained above in the simplest cases with a single axion and single exponent in the superpotential, in string theory the axion decay constants are not large enough to provide inflation and, in particular, inflation with detectable gravity waves. Therefore one would like to use the ideas of assisted inflation \cite{Liddle:1998jc}  in the context of string theory as proposed in the N-flation model in \cite{Dimopoulos:2005ac}.

The main idea of assisted inflation \cite{Liddle:1998jc} is that each of $N$ fields feels the downward force from its own potential, but also the collective Hubble friction from the energy density of all fields. Therefore, slow-roll is easier to achieve and, in addition, one can attain super-Planckian excursions in field space even if the individual fields each have sub-Planckian displacements.

The equations of motion for a set of scalar fields in a generic situation with a moduli space metric (in real notation with $L_{kin}={1\over2}G_{ij}(\phi)\partial_\mu\phi^i\partial^\mu\phi^j$) are \cite{Gaillard:1995az,Sasaki:1995aw}:
\begin{equation}
\ddot \phi^i + \Gamma^i_{jk}(\phi) \dot \phi^j \dot \phi^k + 3 H
\dot{\phi}^i +G^{ij}(\phi)\partial_j V=0 \ . \label{modulispace}
\end{equation}
Here $G^{ij}(\phi)$ is the inverse metric of the moduli space and $\Gamma^i_{jk}(\phi)$ are the Christoffel symbols in the moduli space. If $G_{ij}=\delta_{ij}$ and $V=\sum_i V_i(\phi_i)$, i.e., if the metric of the moduli space is flat and if the potential is a sum of the potentials of the individual fields, the assistance effect becomes clear:
\begin{equation}
\ddot\phi^i+3H\dot{\phi}^i+\partial_i V_i=0\ , \qquad H^2= {\sum_i V_i\over 3 M_{Pl}^2} \ .
\label{flat}
\end{equation}
Each field responds to its own potential (there is no summation in the term $\partial_i V_i$ above), but the friction via the Hubble parameter comes from all of the fields and can be significantly stronger than in case without assistance.

It is clear from the more complicated general formula (\ref{modulispace}) that having $N$ scalars or axions is far from sufficient to guarantee an
assistance effect.  One must also either have good reasons to approximate $V$ as a sum of potentials for individual fields (which can be justified for axions in some circumstances), or one must display the equivalent $1/N$ scaling of slow-roll parameters in the full system (\ref{modulispace}). Indeed, many proposed models of assisted inflation are not radiatively stable and would not be expected, in the end, to enjoy the $1/N$ suppression of slow roll parameters.

One attempt to design a radiatively stable model of assisted large-field inflation that can occur in string theory appears in \cite{Dimopoulos:2005ac}. The model requires a large number of axions,  $ N\sim  240  \left ({M_{Pl}\over f}\right )^2 $, where $f$ is the generic axion decay constant. For $f\approx 10^{-1}M_{PL}$, one should have $N\approx 10^4$. String theory may provide such a large number, there are known compactifications with up to $10^5$ axions. However, \cite{Dimopoulos:2005ac} also described a new, generic concern about the idea of assisted inflation. In a theory with $N$ species and UV cutoff $\Lambda_{UV}$, there is a renormalization of Newton's constant
\begin{equation}
\label{renorm}
\delta M_{Pl}^2 \sim N \Lambda_{UV}^2~.
\end{equation}
Since $M_{Pl}^2$ appears explicitly in the standard slow-roll parameters $\epsilon,\eta$, this $N$ dependence naively ${\it cancels}$ the gain one would have from assistance at parametrically large $N$.  Therefore, it is a UV sensitive question whether the idea works at all, and relies on cancelations at the scale $\Lambda_{UV}$ to produce a small coefficient in (\ref{renorm}).  Circumstances where this cancelation might occur (in the context of the leading $N$-dependent correction in heterotic string models) were described in \cite{Dimopoulos:2005ac,Easther:2005zr}. We shall simply proceed without accounting for this subtlety; if we cannot make a model before including this effect, it will certainly only make matters worse to include it. Further issues with the phenomenology of N-flation (specifically reheating) are discussed in \cite{Green:2007gs}.

Even ignoring this subtlety, one may wonder whether the phenomenological assumptions made in \cite{Dimopoulos:2005ac,Easther:2005zr} can be realized in concrete string models. We will focus here on the issue of whether it is reasonable to treat the dynamics of the axions as inflatons, while ignoring other moduli fields in the problem. The original papers discuss models where axions are partnered with K\"ahler moduli. However, depending on the precise string theory and choice of compact manifold, one can also find models where axions are partnered with themselves. In addition, there
are many different possible ways in which the moduli could be stabilized: the effective potential is derived from a K\"{a}hler potential and a superpotential, either can provide the scale which stabilizes the moduli. If the former is used, the superpotential will still be needed to set of the mass of the axion (necessary for inflation to end), as the K\"{a}hler potential is independent of this component of the modulus. However, K\"{a}hler stabilized models\footnote{This class of models was suggested to us by S. Kachru.} should stabilize the volume modulus, even with a constant superpotential.

We examine superpotential stabilization of the moduli in simple, KKLT-type scenarios in the next section, where we work with a scale free logarithmic K\"{a}hler potential. Following this we then attempt K\"{a}hler stabilization in related models, by considering corrections to the tree level K\"{a}hler potential -- we find that K\"{a}hler stabilization fails in this scenario, the moduli are not stabilized with only a constant superpotential. In section \ref{othermodels} we consider stringy inspired models that do not fit readily into our K\"{a}hler-stabilized/superpotential-stabilized classification -- the superpotential is essential for stabilization, but the K\"{a}hler potential also contributes to the mass scales of the moduli. Section 6 contains supergravity examples where the K\"{a}hler potential participates significantly in the stabilization of the axion partner. Then, the superpotential stabilizes the axion (with a small effect on the stabilization of the partner). In this way we get a significant mass hierarchy between the axion and the partner, which leads to axion inflation.

\section{KKLT-Type Models: Superpotential stabilization}
\label{onemodulus}
We begin our survey with a study of axionic inflation in type IIB string theory compactification on a Calabi-Yau threefold $X$. We assume that the axion-dilaton modulus and complex structure moduli have been stabilized by fluxes and we are only left with $n=h^{1,1}(X)$ K\"{a}hler moduli which need to be stabilized.\footnote{We are really working in an ${\cal N}=1$ supersymmetric orientifold of the Calabi-Yau, and making the (inessential) assumption that all K\"ahler modes were projected in by the orientifold action.} The potential of the four dimensional ${\mathcal{N}}=1$ supergravity is then given by
\begin{eqnarray}\label{potential}
V=e^{K}\Big(\sum_{i=1}^{n}|D_{i}W|^{2}-3|W|^{2}\Big)\ ,
\end{eqnarray}
where the K\"{a}hler potential is given by
\begin{eqnarray}\label{kahlerpot}
K(\{T_{i},\bar{T}_{i}\})=-2\ln\Big[\hat{\mathcal{V}}(T_{i},\bar{T}_{i})\Big]\
,
\end{eqnarray}
where $\hat{\mathcal{V}}$ is the volume of the Calabi-Yau internal space and for the case of one modulus is given by $\hat{\mathcal{V}}=(T+\bar{T})^{3/2}$ \footnote{\ The dependence of $\hat{\mathcal{V}}$ for the general case on $T_{i}$ and $\bar{T}_{i}$ is described in the appendix.}. The superpotential is an analytic function of the K\"{a}hler moduli and has the following general form
\begin{eqnarray}\label{superpot}
W(\{T_{i}\})=W_{0}+W_{1}(\{T_{i}\})\ ,
\end{eqnarray}
where $W_{0}$ refers to the flux superpotential (depending on the complex structure moduli only) and $W_{1}$ indicates the dependence on the K\"{a}hler moduli. The models we consider will mostly have an exponential form for $W_{1}$ -- such a term can arise from non-perturbative effects such as instantons or gaugino condensation and was used in \cite{Kachru:2003aw} to construct meta-stable de Sitter vacua.

A detailed analysis of the above can be found in the appendix, where we demonstrate that at the supersymmetric minima the mass matrix for the K\"{a}hler moduli is given by:
\begin{eqnarray}
&&D_{i}D_{j}V=-e^{K}\bar{W}D_{i}D_{j}W\ ,\label{susyDDV1}\\
&&\bar{D}_{\bar{i}}D_{j}V=e^{K}\Big((D_{j}D_{k}W)(\bar{D}_{\bar{i}}\bar{D}^{k}\bar{W})-2G_{\bar{i}j}|W|^{2}\Big)\
.\label{susyDDV2}
\end{eqnarray}

It has been shown in \cite{Kallosh:2007ig} using  numerical examples that the volume-axion potential in KKLT models with one exponent has the shape of a funnel: in all the studied examples the axion direction was as steep as the volume modulus direction. Moreover, the uplifting procedure does not significantly change the shape of the two-dimensional potential near the minimum. Here we will study this issue analytically for generic parameters of the KKLT model. We will compute the volume-axion mass formula at the generic supersymmetric minimum of the potential and compare the corresponding curvatures of the potential in the volume and axion direction. We will find a very simple formula for the ratio of the axion-volume masses which will be easy to analyze for all available parameters of the potential.

Before immersing ourselves into the particular details of model-building and constraints, there is one further point we should note. In the following three subsections we consider models with a single modulus (and thus a single axion). Needless to say this is not sufficient for N-flation, should we find that the volume and axion decouple, we would need to assemble many such moduli together to obtain IGW-generating inflation. However, given that without the decoupling of the axion and its partner we expect all the fields in the potential to be cross-coupled (see above), the existence of a mass hierarchy between the two components of the modulus appears to be a necessary condition for N-flation in this class of models. It is not inconceivable, though, that the cross-coupled potential will also N-flate, but this situation is not tractable at this time.

As mentioned above, the details of the calculations for this section can be found in the appendix.

\subsection{KKLT: an exact volume-axion mass ratio}
In the KKLT model \cite{Kachru:2003aw}, the \K\ \, potential and the superpotential are given by
\begin{eqnarray}\label{kkltsuppot}
K(T, \bar T) = -3\ln (T+\bar T) \ , \qquad W(T)=W_{0}+Ae^{-aT}\ ,
\end{eqnarray}
where $A$ and $a$ are some (real) constants and we keep $W_0$ complex.
\begin{equation}
T=\sigma+i\alpha\ .
\end{equation}
A little work gives:
\begin{eqnarray}\label{a1}
V=\frac{(aA)^{2}}{6\sigma}e^{-2a\sigma}+\frac{aA^{2}}{2\sigma^{2}}
e^{-2a\sigma}+\frac{aA}{4\sigma^{2}}e^{-a\sigma}
(W_{0}e^{ia\alpha}+\bar{W}_{0}e^{-ia\alpha})\ .
\end{eqnarray}
The masses of the $\sigma$ and $\alpha$ fields will then correspond to the eigenvalues of the following 2-dimensional symmetric matrix:
\begin{eqnarray}\label{d2V}
H=\left(
    \begin{array}{cc}
      \partial_{\sigma}\partial_{\sigma}V & \partial_{\sigma}\partial_{\alpha}V \\
      \partial_{\alpha}\partial_{\sigma}V & \partial_{\alpha}\partial_{\alpha}V \\
    \end{array}
  \right)\ .
\end{eqnarray}
At the supersymmetric minimum the field values are:
\begin{eqnarray}\label{sigalp2}
e^{-2a\sigma_{0}}=\frac{9|W_{0}|^{2}}{A^{2}(3+2a\sigma_{0})^{2}}\
.
\end{eqnarray}
And for the $\alpha$ field:
\begin{eqnarray}\label{sigalp3}
\alpha_{0}=-\frac{\theta}{a}+\frac{n\pi}{a}\ ,\
n\in{\mathbb{Z}}\ ,
\end{eqnarray}
where $\theta$ is defined as the phase of the flux superpotential $e^{i\theta}=\frac{W_{0}}{|W_{0}|}$. The off diagonal elements in the mass matrix $H$ vanish, leaving
\begin{eqnarray}\label{mass}
m_{\sigma}^{2}=\sigma_{0}^{2}H_{11}\ , \ \
m_{\alpha}^{2}=\sigma_{0}^{2}H_{22}\ .
\end{eqnarray}
Note that in order to get the canonical masses we have multiplied by a factor of $\sigma_{0}^{2}$. The masses are:
\begin{eqnarray}
&&
m_{\sigma}^{2}=\frac{3|W_{0}|^{2}}{8\sigma_{0}^{3}}\frac{4(a\sigma_{0})^{2}(2+5(a\sigma_{0})+2(a\sigma_{0})^{2})}
{(3+2a\sigma_{0})^{2}}\ ,\label{masssigma}\\
&&
m_{\alpha}^{2}=\frac{3|W_{0}|^{2}}{8\sigma_{0}^{3}}\frac{4(a\sigma_{0})^{3}}{3+2a\sigma_{0}}\
.\label{massalpha}
\end{eqnarray}
Note that all masses are independent of the value of $\alpha_{0}$ and only depend on the value of $\sigma_{0}$ and $|W_{0}|$. The exact ratio of the masses at the supersymmetric minimum is given by
\begin{eqnarray}\label{ratio}
\frac{m_{\sigma}^{2}}{m_{\alpha}^{2}}=1+\frac{2(1+a\sigma_{0})}{(a\sigma_{0})(3+2a\sigma_{0})}\sim1+\frac{1}{a\sigma_0}\
,
\end{eqnarray}
where the approximation is in the large $a\sigma_{0}$ limit. Noting the form of the kinetic term for the axions:
\begin{equation}
\frac{3}{4\sigma^2}\dot{\alpha}^2\ ,
\end{equation}
it is clear that after rescaling the fields to get a canonically-normalized kinetic term, we obtain:
\begin{equation}
\sqrt{\frac{3}{2}}\frac{1}{\sigma_0}\alpha=\tilde{\alpha}\ .
\end{equation}
Similar arguments hold for the volume kinetic terms, and when the appropriate changes are made in the potential, we find that the effective value of $a$ changes: $\tilde{a}=\sqrt{2/3}\sigma_0a$. Since the effective axion decay constant is given by $f=1/\tilde{a}\sim1/a\sigma_0$, it is apparent that a large mass hierarchy between the axion and its partner is only possible with a large axion decay constant. However, as shown in \cite{Banks:2003sx} (and reviewed above) such decay constants are not available in this regime of string theory. Thus the mass ratio never deviates greatly from 1 and it is clear that KKLT is not a suitable setting for N-flation.

\subsection{KL racetrack models}
The first obvious generalization of the KKLT scenario is to keep the same K\"{a}hler potential and add an extra exponential term, so that the superpotential is of racetrack form:
\begin{eqnarray}\label{klsuppot}
K(T, \bar T) = -3\ln (T+\bar T) \ , \qquad W(T)=W_{0}+Ae^{-aT}+Be^{-bT}\ ,
\end{eqnarray}
where $a$, $b$, $A$, and $B$ are real constants. Such models are analyzed in a stringy context in \cite{Kallosh:2004yh,Blanco-Pillado:2005fn}. This expression for $W$ gives the following potential:
\begin{eqnarray}
V=\frac{1}{6\sigma^2}\left(aA^2e^{-2a\sigma}(3+a\sigma)
+bB^2e^{-2b\sigma}(3+b\sigma)
+ABe^{-(a+b)\sigma}(3(a+b)+2ab\sigma)\cos[(a-b)\alpha]\nonumber\right.\\
\left.+\frac{3}{2}aAe^{-a\sigma}(W_{0}e^{ia\alpha}+\bar{W}_{0}e^{-ia\alpha})
+\frac{3}{2}bBe^{-b\sigma}(W_{0}e^{ib\alpha}+\bar{W}_{0}e^{-ib\alpha})\right)
\end{eqnarray}

The minimization conditions for the supersymmetric vacua, $D_{T}W=0$, give:
\begin{eqnarray}
\frac{3\textrm{Re}(W_0)}{2\sigma_0}&=&-Ae^{-a\sigma_0}\left(a+\frac{3}{2\sigma_0}\right)\cos(a\alpha_0)
-Be^{-b\sigma_0}\left(b+\frac{3}{2\sigma_0}\right)\cos(b\alpha_0) \ , \nonumber\\
\frac{3\textrm{Im}(W_0)}{2\sigma_0}&=&Ae^{-a\sigma_0}\left(a+\frac{3}{2\sigma_0}\right)\sin(a\alpha_0)
+Be^{-b\sigma_0}\left(b+\frac{3}{2\sigma_0}\right)\sin(b\alpha_0) \ .
\end{eqnarray}
Evidently, the above do not separate nicely into independent equations for $\sigma$ and $\alpha$ in a straightforward fashion. Accordingly, the mass matrix is somewhat involved and its eigenvalues do not give particular pleasant expressions. However, before leaping into the unpleasant details, a little thought reveals the analysis to be a much simpler task. Treating the term with larger of $a,b$ in the exponent as a perturbation to the KKLT model, it is clear that the size of the correction is of order $e^{-|a-b|}$ -- this is exponentially small unless we tune $|a-b|$ to be close to
zero. As a result, the only chance of finding a valley (given that none exists in KKLT) is with this tuning. With this simplification, it is straightforward to show that to first order in $|a-b|$ the mass ratio is given by (we've chosen $b>a$)
\begin{equation}
\left(1+\frac{2}{3a\sigma_0}+\frac{2}{9+6a\sigma_0}\right)-\frac{2B(3+4a\sigma_0+2a\sigma_0^2}{(A+B)(a\sigma_0)^2(3+2a\sigma)^2}(b-a)\sigma_0\
.
\end{equation}
To first order in $1/a\sigma_0$ and $1/b\sigma_0$ this becomes
\begin{equation}
\frac{m_{\sigma_0}^2}{m_{\alpha_0}^2}\sim1+\frac{1}{a\sigma_0}-\frac{B}{A+B}(b-a)\sigma_0\left(\frac{1}{a\sigma_0}\right)^2\
.
\end{equation}

Given that we expect $A$ and $B$ to be $O(1)$ numbers, once again we have failed to find a large mass hierarchy -- once again we use the result of \cite{Banks:2003sx} for the axion decay constant of the canonically-normalized field. With the benefit of hindsight, this is not at all surprising. The additional term in the KL model is exponentially small (thus we neglect it and get KKLT) or
is tuned to be almost exactly the same as the original non-perturbative effect (so the two terms can be combined, also
giving KKLT).

\subsection{More Instantons}
The models above can be extended to superpotentials with more exponential terms. Basically, any additional instanton can only have an effect if the coefficients in the exponent ($a_i$) are almost equal. If this isn't the case then there will be some subset of terms that are exponentially suppressed -- reducing the model to a simpler one. However, if we tune the $a_i$ we're stuck for the reasons outlined above. The superpotential can be written as (choosing $a_1$ to be the smallest of the $a_i$):
\begin{eqnarray}
W&=&W_0+\sum_{i=1}^NA_ie^{-a_iT}=W_0+A_1e^{-a_1T}\sum_{i=1}^NA_ie^{-(a_i-a_1)T}\
.
\end{eqnarray}

Unfortunately finding the mass ratio analytically is not straightforward. However, it is reasonable to expect that the two exponent case generalizes.

\subsection{Better racetrack inflationary models}
A numerical example of stabilization of two complex moduli is given by a ``better racetrack'' inflation model \cite{Blanco-Pillado:2006he}, related to IIB CY flux compactification on the orientifold $\mathbb{P}^4_{1,1,1,6,9}$. The K\"{a}hler potential for these two moduli is
\begin{equation}
K= -2\ln
{\cal V} \ , \qquad {\cal V}=(T_1+\bar T_1)^{3/2}- (T_2+\bar
T_2)^{3/2}
\end{equation}
and the superpotential
\begin{equation}
W=W_0 + Ae^{-aT_1} +Be^{-bT_2}
\end{equation}

With some fine-tuning one can produce a saddle point in the potential with a very flat direction in a combination of the axion moduli direction which leads to a modular inflation compatible with observations. When the system is kicked out from the top of the potential it enters into a waterfall regime and quickly reaches the minimum of the potential. This model predicts no IGW.

Here we would like to examine the relation between the curvatures of the axions and the corresponding volumes near the minimum of the potential. The only difference from the KKLT model which we studied before is that we have two moduli instead of one. Now it is more difficult to handle the issue analytically. However a numerical example may be useful, it will either point to a possibility to get away from the constraints of the single KKLT model or will confirm the problem.

We find the following eigenvalues for the masses of these two complex moduli, $T_1= \sigma_1+ i a_1$ and $T_2= \sigma_2+ i a_2$
\begin{equation}
m_{\sigma_1}= 1.10\times 10^{-16} \ , \quad m_{a_1}= 1.03\times
10^{-16} \ ; \qquad m_{\sigma_2}= 2.13\times 10^{-18} \ , \quad
m_{a_2}= 1.52\times 10^{-18}
\end{equation}
A quick glance at these masses shows that, as predicted in KKLT models with one modulus, in the case of two moduli we find the volume directions near the minimum are of the same steepness as the axion directions for each of the two moduli. This means that, again, we do not get the desired axion valley.

\section{KKLT-Type Models: Attempts at K\"{a}hler Stabilization}\label{kahlerstab}
So far we have shown that, in the string theory models we have considered, the axion and its partner have masses of approximately the same order. This is principally because both fields are simultaneously stabilized and thus the masses have the same dependence on the parameters of the model.

With the above observations in mind we wish, in this section, to use the idea of K\"{a}hler stabilization to separate the stabilization process of the axion and its partner (which we call ``the volume field'' as in this section the relevant moduli will be the volumes of the 2-cycles and 4-cycles). We consider K\"{a}hler potentials that only depend on the volume moduli and try to stabilize this moduli in the absence of any superpotential dependence. To do this we will first study shift-symmetric K\"{a}hler potentials and neglect the non-perturbative instanton/gaugino condensation corrections to the constant superpotential. This will leave flat axion directions. We find below that in this case theoretical conditions make it impossible to construct an axion valley.

\subsection{Shift-symmetric K\"{a}hler potentials in string theory}
We begin by summarizing what is considered to be the general class of computable K\"{a}hler potentials associated with Calabi-Yau compactifications. The first  studies in \cite{Candelas:1990pi} - \cite{Hosono:1994av} were performed in early 90s before the second string revolution and were based on Calabi-Yau compactifications in the context of the two-dimensional $\sigma$-model
\begin{equation}
S= {1\over 2\pi \alpha'} \int d^2\sigma \sqrt{h}(h^{\alpha \beta}
G_{ij}(\phi)+\epsilon^{\alpha \beta}B_{ij})\partial_\alpha \phi^i
\partial_\beta \phi^j +... \ .
\end{equation}
About ten years later there was a renewal of interest in computable K\"{a}hler potentials, but now in the context of flux compactification with non-perturbative Ramond-Ramond forms. The computation was performed in \cite{Becker:2002nn} in the context of dimensional reduction of the effective ten-dimensional type IIB supergravity, taking into account stringy $\alpha'$ corrections \footnote{More recently an attempt was made in \cite{Berg:2007wt}  to use the string loop corrections to K\"{a}hler potential, in addition to $\alpha'$-corrections,  for the purpose of volume stabilization without non-perturbative corrections to the superpotential. The authors find that these attempts so far were either not satisfactory
or inconclusive. Therefore we will not use these models in our current analysis. However, if this class of models will be developed and improved, it may become  useful for our scenario.}. The relevant K\"{a}hler potentials are closely related to those known from previous $\sigma$-model studies and have been used for flux compactification and moduli stabilization -- starting with \cite{Denef:2004dm} and \cite{Balasubramanian:2004uy,Bobkov:2004cy,Cicoli:2007xp}; see also the review in \cite{Douglas:2006es}.

Since our goal is not to miss any corner of the stringy landscape with interesting cosmology we will show no bias towards one or other of these K\"{a}hler potentials. From a phenomenological point of view our only concern is that the complex structure moduli are stabilized by some mechanism and so do not interfere with the inflationary dynamics.

We start with early papers on CY compactification, summarized in the TASI lectures \cite{Hosono:1994av}. These models in string theory are related to ${\cal N}=2$ supergravity, which is described by the prepotential ${\cal F}(X^I)$ depending on the holomorphic homogeneous coordinates $X^I= (X^0, X^i)$. The prepotential is homogeneous of degree 2 and can be also represented as a function of special coordinates $t^i= {X^i\over X^0}$ so that ${\cal F}(X^0,
X^i)= (X^{0})^2 F(t^i)$. Each of the moduli has as a real part an axion $a^i$ originating from a form field and also a positive imaginary part $v^i$:
\begin{equation}
t^i= a^i+ i v^i \label{moduli} \ .
\end{equation}
The relation between the K\"{a}hler potential and the prepotential in the $X^0=1$ gauge is
\begin{eqnarray}\label{prekah}
K(t^{i},\bar{t}^{i})=-\ln\Big[ i
\Big(2({{F}}-\bar{{F}})-(t^{i}-\bar{t}^{i})\Big(
\frac{\partial{{F}}}{\partial
t^{i}}+\frac{\partial\bar{{F}}}{\partial\bar{t}^{i}}\Big)\Big)\Big]\ .
\end{eqnarray}
This formula for the K\"{a}hler potential is equivalent to the original one given in \cite{deWit:1984pk}, where special geometry was introduced:
\begin{equation}
K= -\ln (i\bar X^I {\cal F}_I - i X^I \bar {\cal F}_I) \ .
\end{equation}
In the limit of  large $v^i $ the prepotential is \cite{Hosono:1994av}
\begin{equation}
F(t)= {1\over 3!} c_{ijk}t^i t^j t^k + {1\over 2} c_{ij}t^i t^j +
c_i t^i +c + F_{inst} \ . \label{large}
\end{equation}
Here $c_{ijk}$ are real Yukawa couplings defined by the classical intersection numbers, which are computable for particular Calabi-Yau manifolds. It is easy to check that the real parts of the constants $c_{ij}, c_i, c$ drop from the K\"{a}hler potential. On the other hand, the presence of $\rm Im \, c_{ij} $ and $\rm Im \, c_i $ terms would break the continuous shift symmetry of the K\"{a}hler potential under
\begin{equation}
t^i\rightarrow t^i+\alpha^i\ , \qquad \alpha^i \quad \rm real
\end{equation}
The instanton part of the prepotential depends on $q^i=e^{2\pi i t^i}$
\begin{equation}
F_{inst}(q^i)\ , \qquad q^i=e^{2\pi i t^i}=e^{-2\pi v^i} e^{2\pi i
a^i}
\end{equation}
Thus $F_{inst}$ breaks the continuous shift symmetry down to a discrete one and in string theory it is believed to be the only source of breaking of the continuous axion shift symmetry. This is based on the fact that the underlying $\sigma$-model has such a symmetry before the world-sheet instantons are taking into account. We may either accept this rather standard assumption of string theory or simply argue that for our search for an axion valley in the landscape we would like to start with a shift symmetric K\"{a}hler potentials. As such we require, as in \cite{Hosono:1994av}, that
\begin{equation}
\mbox{Im} \, c_{ij} =0, \qquad \mbox{Im} \, c_i =0 \ ,
\end{equation}
and we will ignore the instanton part as it also breaks the continuous shift symmetry of the K\"{a}hler potential. This leaves us with
\begin{equation}
K=-\ln\left [i\left( {1\over 3!}c_{ijk}(t-\bar t)^i (t-\bar t)^j
(t-\bar t)^k + 2(c-\bar c)\right)\right ]\ . \label{cubic+c}
\end{equation}
Taking into account our definition of the special coordinate $t^i$ in (\ref{moduli}) and the known $\alpha'$ corrected value of the imaginary part of $c$, we can relate it to the Euler number of the Calabi-Yau manifold $\chi$. Rewriting $K$, we obtain, up to a constant factor:
\begin{equation}
K= -\ln [ {\cal V}  + {\xi\over 2}]\ , \qquad {\cal V}={1 \over
3!}c_{ijk} v^i v^j v^k \ , \qquad \xi=-{\chi \zeta(3)\over 4
(2\pi)^3} \ . \label{K}
\end{equation}
This expression for $K$ is deduced from \cite{Hosono:1994av} and it is based on $\alpha'=1$ units used there.

Now, we move onto $\alpha'$ corrections in the presence of fluxes, explored recently in the context of type IIB string theory (with fluxes) \cite{Denef:2004dm, Becker:2002nn, Balasubramanian:2004uy, Douglas:2006es}. The corrections to the K\"{a}hler potential can be presented as follows:
\begin{equation}
K=-\ln \rm Im \tau- \ln \int_Z \Omega(z) \wedge\bar \Omega(\bar z) -
2\ln \left [\hat {\cal V}  + {\hat \xi\over 2}\right ]\ . \label{K1}
\end{equation}

There is a combined axion-dilaton  and Calabi-Yau moduli space, along with a shape moduli space ${\cal M}_c(Z)$ and the size moduli space ${\cal M}_k(Z)$. The axion-dilaton is a field $\tau= a+ie^{-\phi}$, $\Omega$ is the holomorphic 3-form on $Z$ and $\hat {\cal V}$ is the volume of the CY space. The K\"{a}hler potential in this form has to be treated as a function of the 4-volume cycles defined via dual coordinates
\begin{equation}
\tau_i\equiv\partial_{v_i}{\cal V}= {1\over 2} c_{ijk}  v^j v^k \ ,
\qquad {\cal V}\equiv { 1 \over 3!}c_{ijk} v^i v^j v^k
\end{equation}
and the ``hat'' induces some dependence on the dilaton due to transformation to the Einstein frame in effective 4d supergravity
\begin{equation}
\hat v^i= v^i e^{-\phi/2} \ , \qquad \hat {\cal V}= {\cal
V}e^{-3\phi/2}
\end{equation}
The dual coordinates $\hat \tau_i$ are combined with form fields to form complex coordinates associated with the volumes of the 4-cycles,
\begin{equation}
\rho_i = b_i+ i \hat \tau_i\ .
\end{equation}
Therefore the last term in (\ref{K1}) is
\begin{equation}
K(\rho,\bar \rho) = - 2\ln \left [\hat {\cal V} + {\hat \xi\over
2}\right ]= -2 \ln ({\rm Im } \rho_i \; \hat v^i +{\xi\over 2}
e^{-3\phi/2}) \label{K2}
\end{equation}
and $\hat v^i$ has to be treated as a function of $\rho_i, \bar \rho_i$. The $\alpha'$ correction term ${\xi\over 2}$ depends on the Euler number  of the Calabi-Yau manifold $\chi$ as in (\ref{cubic+c}) and also on the string coupling, $\rm Im \tau= e^{-\phi/2}$. The superpotential depends on the axion-dilaton and on the shape moduli and it can be fixed by requiring that $D_\tau W= D_{z}W=0$. In absence of the $\alpha'$ corrections when $\xi=0$, the action is of the no-scale form and the potential vanishes if the superpotential is independent on size moduli \cite{Becker:2002nn}.

\subsection{Stabilization of volume moduli with a constant superpotential}
Both K\"{a}hler potentials discussed above (given by (\ref{K}) and (\ref{K2})) have interesting features which define the stabilization of volume moduli in the presence of a constant superpotential. Interestingly, there is little qualitative difference between the two cases, but for completeness, both cases have been analyzed and the details of the results can be found in appendix C.

The bottom line is that, in cases which volume moduli can get stabilized via the K\"{a}hler stabilization mechanism,  the stabilized volume of the compactification is obtained to be proportional to $\xi$. However, as a physical requirement, these class of models are reliable only in the limit of large compactification volume. Therefore, these class of models are not an honest candidate to stabilize the volume moduli by this mechanism.

\section{$\alpha'$ Corrected K\"{a}hler Potential With a Non-Perturbative Superpotential}\label{othermodels}
As we mentioned above, successful K\"{a}hler stabilization would have to include the effects of non-perturbative corrections to the superpotential. Even though the machinations of the previous section were unsuccessful there are interesting examples in the literature which use the setup of \cite{Balasubramanian:2004uy}, but with a non-constant superpotential. The two examples we give are not K\"{a}hler stabilized (obviously), but in each the K\"{a}hler potential plays an important role in determining the scales of the masses.

\subsection{Large Volume stabilization models}
Here we use the information on the phenomenology of the working model \cite{Conlon:2005ki} originating from the type IIB CY flux compactification on the orientifold $\mathbb{P}^4_{1,1,1,6,9}$, with a $K$ given as in \cite{Balasubramanian:2004uy}. We will analyze the axion and volume moduli masses near the minimum of the potential. The model is valid in approximation of large volume ${\cal V}$. The axion-dilaton and complex structure moduli are fixed by fluxes.  The K\"{a}hler potential of the \K\  moduli $T_5= \tau_5+ib_5$ and $T_4=\tau_4+ib_4$ is given by $K= -2\ln( {\cal V}+{\xi\over 2})$
where
\begin{equation} {\cal V}={1\over 9 \sqrt 2}(\tau_5^{3/2}-
\tau_4^{3/2})
\end{equation}
Here $\tau_5$ is the large volume modulus and $\tau_4$ is the small volume modulus and $\tau_5\gg \tau_4 > 1$ in string units $l_s= 2\pi \sqrt \alpha'$. At the point of stabilization where ${\cal V}= {\cal V}_0$ the masses of the moduli fields are
\begin{equation}
m^2_{\tau_5}= {g_s^2 W_0^2\over 4\pi {\cal V}_0^3}M_{Pl}^2 \ ,
\qquad m^2_{b_5}= e^{- ({\cal V}_0)^{4/3}}M_{Pl}^2
\end{equation}
The ratio of these two fields is given by
\begin{equation}
{m^2_{\tau_5}\over m^2_{b_5}}= {g_s^2 W_0^2 e^{ ({\cal
V}_0)^{4/3}}\over 4\pi {\cal V}_0^3}
\end{equation}
At first glance this looks like an excellent candidate for the axion valley potential. However, a closer look reveals a problem: the axion, in
all cases of interest, is way too light. There are 3 regimes for the supersymmetry breaking parameter in these classes of models. In case of the GUT scale   the masses of the volume and the axion are
\begin{equation}
m_{\tau_5}= 2.2\times 10^{10}GeV \ , \quad m_{b_5}=
10^{-300} GeV \ , \label{1}
\end{equation}
for intermediate scale
\begin{equation}
m_{\tau_5}= 22 GeV \ , \quad m_{b_5}= e^{-10^{6}}
GeV \label{2}
\end{equation}
and for TeV scale
\begin{equation}
m_{\tau_5}= 2.2\times 10^{-26}GeV \ , \quad m_{b_5}= e^{-10^{18}} \ .
GeV \label{3}
\end{equation}
In all cases the curvature in the volume modulus is indeed much larger than the curvature in the axion direction near the minimum, which is what we wanted. However, the mass of the $b_5$ axion is too small, the potential is too flat in the axion direction. The value of the axion mass near
the minimum of the potential which will provide us with the natural inflation can be evaluated from the expression $\Lambda^4(1-\cos(\phi/f))$. Here $\Lambda$ is of the GUT scale and $f\sim 25 M_{Pl}$ to have an agreement with observations. One finds that in the quadratic approximation the mass of the axion should be of the order of $10^{13}$GeV. This is many order of magnitude larger than all values of the axion masses in models of \cite{Conlon:2005ki}.

The masses of the small volume modulus $\tau_4$ is equal to the mass of its axion partner  $b_4$, as in the KKLT model. Therefore the $b_4$ axion is not suitable for the axion valley potential we seek. We have avoided the issue of stabilization only taking place in untenable regions of moduli space; however, as with the constant superpotential, only one moduli gets a hierarchy. This in turn means that even excepting the cosmological problems with the excessively light axion we still would have problems N-flating.

\subsection{\K\ uplifting models}
Another class of moduli stabilization models was proposed in \cite{Westphal:2006tn} where the volume moduli are stabilized directly in  de Sitter minima (instead of AdS minima of KKLT models or large volume stabilization models in \cite{Balasubramanian:2004uy,Conlon:2005ki} which require an
uplifting with anti-D3 branes or fluxes on D7 branes). These models have a K\"{a}hler potential analogous to the one in the previous subsection. It is interesting to present here the masses of the volume-axion moduli, from the examples of \cite{Westphal:2006tn}, where these models were studied in detail.
\begin{itemize}
\item Model with 2 complex moduli, $T_1= \sigma_1+ia_1$ and $T_2= \sigma_2+ia_2$
\begin{equation}
m^2_{\sigma_1}\approx 10^{-5}\ , \quad  m^2_{a_1}\approx 5\times
10^{-6}\ ; \qquad m^2_{\sigma_2}\approx 6\times 10^{-8}\ , \quad
m^2_{a_2}\approx 1.4\times 10^{-7}
\end{equation}
\item Model with 3 complex moduli, $T_1= \sigma_1+ia_1$, $T_2= \sigma_2+ia_2$ and $T_3= \sigma_3+ia_3$
\begin{equation}
m^2_{\sigma_1}\approx 3\times 10^{-5}\ , \quad  m^2_{a_1}\approx
3\times 10^{-5}\ ; \qquad  m^2_{\sigma_2}\approx 3\times 10^{-6}\ ,
\quad m^2_{a_2}\approx  10^{-6}
\end{equation}
and
\begin{equation}
m^2_{\sigma_3}\approx 1.5\times 10^{-8}\ , \quad  m^2_{a_3}\approx
4\times 10^{-8}
\end{equation}
\end{itemize}
One can easily see that for each complex modulus shown above the ratio of the volume modulus mass to the axion mass is of the order 1. Once again we find that in known models with logarithmic K\"{a}hler potential and stabilization of moduli via the superpotential there is no axion valley.

\section{Axion Valley Model}\label{sugramodels}
\subsection{The axion valley model (natural inflation in supergravity)}
The natural inflation  PNGB (Pseudo-Nambu-Goldstone Boson) model \cite{Freese:1990rb} is based on a potential of the form $\Lambda^4(1\pm \cos (\phi/f)]$ with $f\geq 0.7 M_{Pl}^{old}\approx 3.5 M_{Pl}$ and $\Lambda\sim M_{GUT}$. It was shown in \cite{Kallosh:2007ig} how to embed natural inflation in supergravity.

\begin{figure}
\centerline{ \epsfxsize 2.7 in\epsfbox {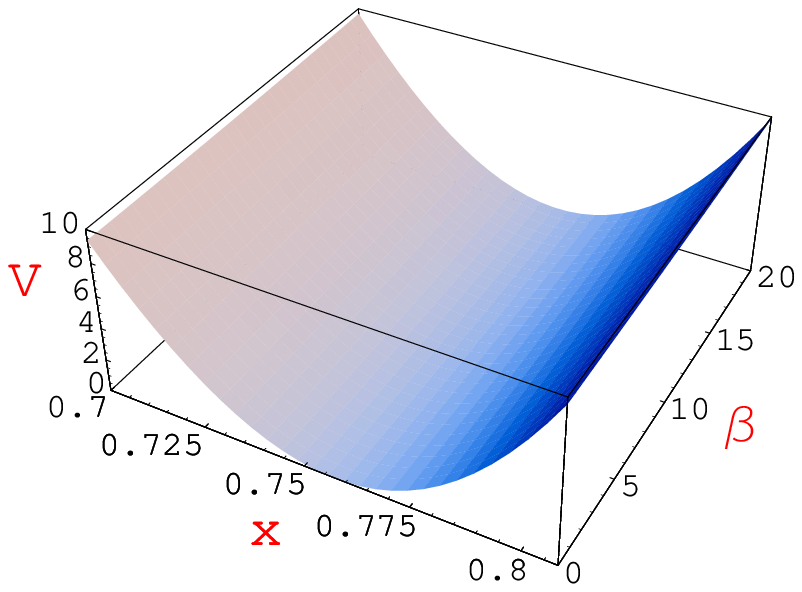} \hskip 0.3cm
\epsfxsize 2.7 in \epsfbox{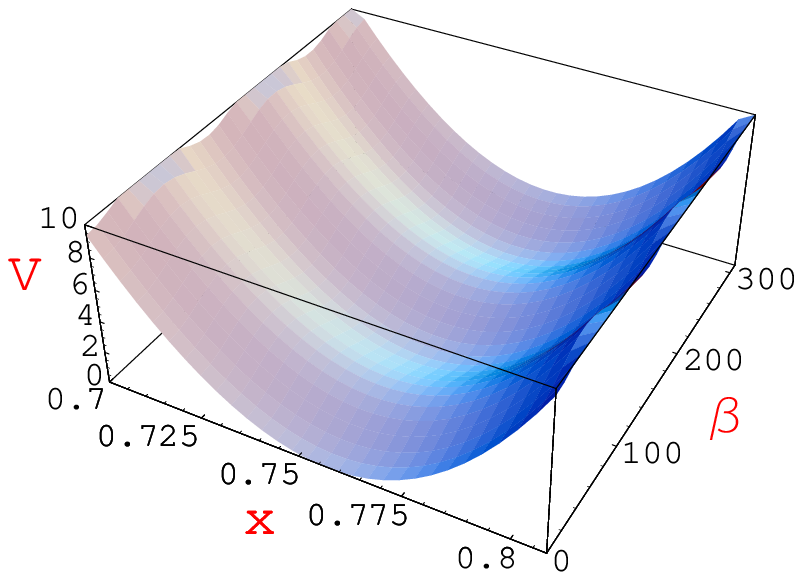}}
\caption{Axion valley potential (\ref{full}), (\ref{valley}). On the left there is a  view of the axion valley. There is a sharp minimum for $x$ and a very shallow minimum for $\beta$. The $\beta$-direction is practically flat for $\beta$ from $0$ to $20$ (in Planck units), whereas  in the $x$-direction the potential appreciates significantly when  $x$ changes by $0.1$. On the right, the potential is plotted for $\beta$ from $0$ to 300. The plot shows the periodicity in  the axion variable $\beta$. Both $\beta$ and $x$ have canonical kinetic terms.} \label{yy2.eps}
\end{figure}

We  consider  the KKLT  model with all fields fixed at their minima, and add to it a field $\Phi$ with a shift-symmetric K\"{a}hler potential and a simple non-perturbative superpotential  which breaks this shift symmetry
\begin{equation}
K={1\over 4} (\Phi+\bar \Phi)^2\ ,  \qquad  W= w_0 + B e^{-b \Phi}\
, \label{axionvalley}
\end{equation}
with
\begin{equation}
V_{\Phi}=
e^K (|DW|^2- 3 |W|^2)= V_1(x) -V_2(x) \cos (b\beta) \ .
\end{equation}
Here $\Phi= x+i\beta$ and
\begin{equation}
V_1(x)=e^{x(-2b+x)}B^2(-3+2( x-b)^2+e^{2bx}(-3+2x^2)w_0^2 \ ,
\label{V1}
\end{equation}
\begin{equation}
V_2=2Be^{bx}w_0(3+2bx-2x^2)  \ . \label{V2}
\end{equation}
The presence of the KKLT model modifies the potential constructed from (\ref{axionvalley}) in two ways. It rescales the overall value of the $\Phi$ field potential and adds to it a positive constant. The effective uplifting can set the potential at the minimum of $\Phi$ close to zero (from the positive side). This rescaling can be absorbed by a rescaling of $w_0$ and $B$. Thus we have a model with canonical kinetic terms for both $x$ and $\beta$ and the following potential
\begin{equation}
g^{-1/2} L= {1\over 2}[ (\partial x)^2+ (\partial \beta)^2] -
V(x,\beta)\ , \label{full}
\end{equation}
where the axion valley potential is
\begin{equation}
V(x,\beta)= V_1(x)- V_2(x) \cos (b\beta)-V_0, \qquad V_0= V_1(x_0)-
V_2(x_0)  \cos (b\beta_0) \ , \label{valley}
\end{equation}
and $x_0,\beta_0$ is the point where the potential both has a minimum and vanishes. $V_1(x), V_2(x)$ are given in equations (\ref{V1}), (\ref{V2}).

\begin{figure}
\centerline{\epsfxsize 3.2 in\epsfbox {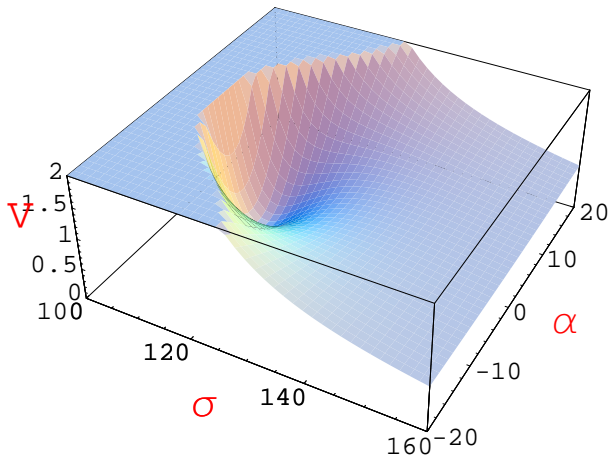}}
\caption{The funnel-type potential of the KKLT model with logarithmic shift symmetric K\"{a}hler potential depending on the volume $\sigma$ and the axion $\alpha$.  The potential in the axion direction is as steep as in the volume modulus direction.} \label{KKLTpot.eps}
\end{figure}

If the minimum is at $\beta_0=0$ the potential at $x=x_0$ takes the form of the natural PNGB model potential \cite{Freese:1990rb,Savage:2006tr}
\begin{equation}
V=V_2(x_0)(1 -\cos(b\beta)) \ . \label{natural}
\end{equation}

Our goal is to make the potential in (\ref{full}) steep for the $x$ field and very flat for the $\beta$ field. This is possible, unlike the KKLT model \cite{Kachru:2003aw}, where the potential is equally steep for the volume modulus and the axion near the minimum of the potential, see figure \ref{KKLTpot.eps}.

In the axion valley model, for parameters $B=1$, $b=0.05$ and $w_0=10^{-4}$, we find that the potential in the $x$ direction is steep, while we have a nearly flat valley for the axion $\beta$. Thus the latter may play the role of the inflaton field -- see figure \ref{yy2.eps}. Note, the minimum of the potential at $\beta=0$ and $x\sim 0.76$ is not supersymmetric.

To make this model compatible with the WMAP3 data, we put the system at the minimum $x=x_0$ and use the values for the parameters suggested in \cite{Savage:2006tr} for the potential $V=\Lambda^4(1-\cos(\phi/f)$. We need  $V_2(x_0)= \Lambda^4$ with $\Lambda$ at the GUT scale and our parameter $1/b$ corresponds to $\sqrt {8\pi}f$ in \cite{Savage:2006tr}. We have to take into account that in supergravity setting we are working in units where
$M_{Pl}=2.4\times10^{18}$GeV=1.

There are two limiting cases to consider. In the first case, $3.5\leq f\ll 25 $ ($0.04\ll b\leq 0.28$), inflation takes place near the maximum of the potential, as in the  new inflation scenario. In the second case,  $f\geq 25$ ($b\leq 0.04$), the potential is very flat at the minimum and the model is close to the simplest chaotic inflation scenario.  In this regime, for $x_0<1$,   the COBE/WMAP normalization of inflationary perturbations implies that $ w_0 B\, b^2 \sim 1.5\times 10^{-12}$. Clearly, such parameters are possible from the point of view of supergravity, particularly taking account of the rescaling mentioned above.

\subsection{Keeping multi-instanton corrections small}\label{sugramulti}
The model above has a problem of the kind discussed in \cite{Banks:2003sx} and reviewed above in section \ref{Banks}. If the source of the exponential term in the superpotential is from instantons, one has to keep the argument in the real part of the exponent $e^{-b \Phi}$ not small, which means
\begin{equation}
b x_{0}> 1 \label{largeexp}
\end{equation}
where $x_0$ is the critical value of $x= \rm Re \, \Phi$. However, in the numerical example above we find that the condition (\ref{largeexp}) is not met when we satisfy the experimental requirements.

\begin{figure}
\label{doublevalley}
\centerline{ \epsfxsize 2.7 in \epsfbox {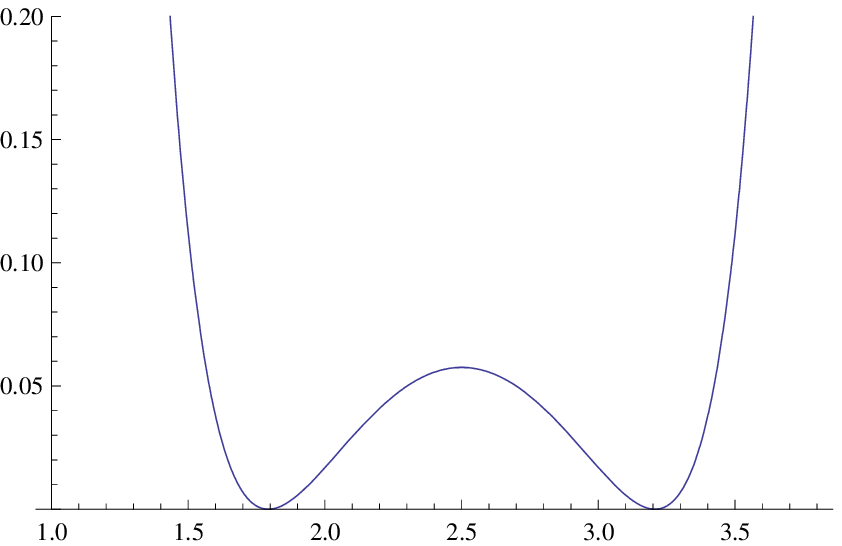} \hskip 0.3cm
\epsfxsize 2.7 in \epsfbox  {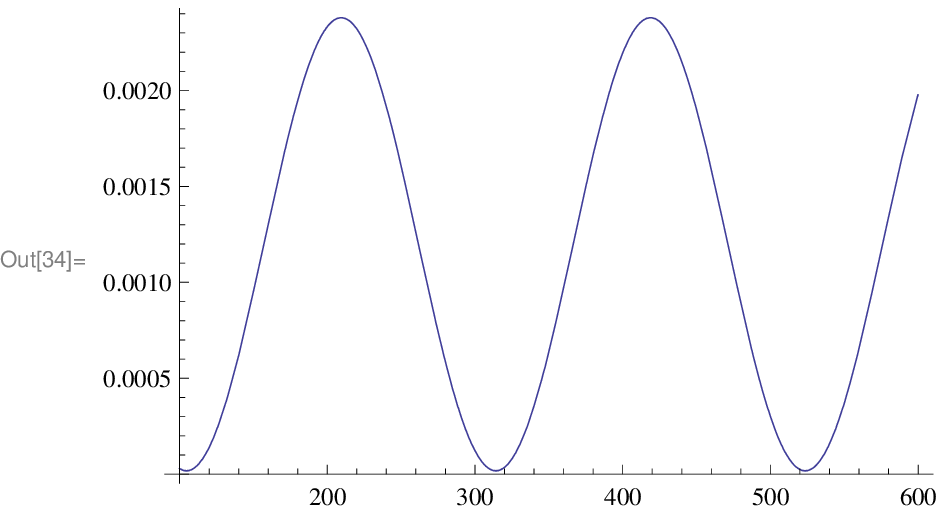}}
\caption{Volume-axion slices of the double valley potential (\ref{U}). On the left the volume slice of the potential is plotted at the minimum for the axion at $\beta={100\pi\over 3}$. The plot shows two minima and a maximum in between. On the right there is a slice of the axion periodic potential with the period ${2\pi\over a-b}$ at the minimum of the volume, $x= 3.21$ in our example, with $(a-b)=3/100$. There is a maximum at $\beta= 2n{100\pi\over 3}$  and a minimum at $\beta= (2n+1){100\pi\over 3}$, $n=0,1, 2, ...$. } \label{Radion.eps}
\end{figure}

Here we will propose a simple generalization of the model with one exponent from the previous section: we will use the racetrack example with two exponents. A qualitative discussion of the situation follows from the features of the potential in this case. We take
\begin{equation}
K={1\over 4} (\Phi+\bar \Phi)^2\ ,  \qquad  W= W_0 + A e^{-a \Phi}+
B e^{-b \Phi}\ . \label{axionvalley2}
\end{equation}
To illustrate why the second exponent allows us to fix the problem, consider the limiting case $W_0=0$. The potential $U(x, \beta)=
e^{K}\Big(\sum_{i=1}^{n}|D_{i}W|^{2}-3|W|^{2}\Big)- U_0$ is
\begin{equation}
U(x, \beta) = U_1(x)+ U_2(x)\cos [(a-b)\beta]-U_0 \ , \label{U}
\end{equation}
where
\begin{equation}
 U_0=  U_1(x_0)+ U_2(x_0)\cos [(a-b)\beta_0] \ .
\end{equation}
Here
\begin{equation}
U_1(x)= e^{x(x-2a-2b)}\left (A^2 e^{2bx}(-3+2a^2-4ax+2x^2)+B^2
e^{2ax}(-3+2b^2-4bx+2x^2)\right)
\end{equation}
and
\begin{equation}
U_2(x)= e^{x(x-a-b)}\left (2 AB (-3+2ab-2(a+b) x+2x^2)\right) \ .
\end{equation}
The axion period is defined by the inverse of the difference between the arguments in the exponents, $(a-b)$ instead of the argument $b$ as in (\ref{valley}). Thus we need to make $(a-b)$ small without having small each of $ax_0$ and $bx_0$. A numerical example of such a situation is given by $a = 2.5,\ b = 2.53,\ A=10,\ B=-0.01$. In figure \ref{doublevalley} we show the slices of this potential in the radial as well as axion directions. There are two axion valleys, one at $x\approx 3.2$ and the other at $x\approx 1.8$, and a maximum in the radial direction in between them. This maximum is supersymmetric before the uplifting, $DW=0$, and the axion direction is flat.

The axion valley model provides the first explicit realization of the natural inflation in supergravity. It realizes the standard lore that the shift symmetry of the K\"{a}hler potential may protect a nearly flat axion potential. It gives a simple  example of such a model, where  the partner of the axion is stabilized and the total potential has a stable  non-supersymmetric minimum. To make this minimum dS we have added a constant to the potential which comes from the KKLT type uplifting procedure with a fixed volume modulus.

\subsection{On a possibility of shift-symmetric quadratic K\"{a}hler potentials in string theory}
To justify the shift-symmetric quadratic K\"{a}hler potentials in string theory one has to be able to argue that the corrections terms  to
the K\"{a}hler potential, for example of the form  $\alpha (\phi + \bar\phi)^4$ and higher order terms will not change the dynamics of the quadratic model significantly. Indeed in cases which we will describe below such higher order terms will naturally appear in string theoretic models from the expansions of the  logarithmic potentials with regard to some volume moduli of small cycles.

A discussion of the possibility of justifying the shift-symmetric quadratic K\"{a}hler potentials in string theory was given in \cite{Kallosh:2007ig}.  These types of K\"ahler potentials can be identified, e.g. in the D3/D7 model, via an expansion of the logarithmic potential, see equations (18,19) in \cite{Kallosh:2007ig}.

\subsubsection{ Axions in $K3\times {T^2\over \mathbb{Z}_2}$}
Here we  describe an analogous model (with regard to shift-symmetric approximately quadratic K\"ahler potential)  in type IIB string theory with a
$K3\times {T^2\over \mathbb{Z}_2}$ compactification, without moving branes. The M-theory version of this model for compactification on $K3\times K3$ was studied in \cite{Aspinwall:2005ad}. In \cite{Tripathy:2002qw} and \cite{Andrianopoli:2003jf} the stabilization by fluxes of part of moduli space of the $K3\times{T^2\over \mathbb{Z}_2}$ compactification was performed. All remaining moduli in this model were stabilized by the instanton corrections to the superpotential in \cite{Aspinwall:2005ad,Bergshoeff:2005yp}. The K\"ahler potential of the part of the moduli space stabilized by instantons is given by the logarithm of the  cubic polynomial
\begin{equation}
K=- \ln \left( (T+\bar T) [(x_0+\bar x_0)^2-\sum_{i=1}^{i=19}(x_i+\bar x_i)^2]\right) \ .
\end{equation}
Here $T+\bar T$ is the K\"ahler modulus of the K3 manifold, and $x_0+\bar x_0$ is the K\"ahler modulus of the torus. The K\"ahler moduli $x_i+\bar x_i,\,  i=1, ..., 19$ are representative of the coset space $S0(3,19)\over SO(3)\times SO(19)$ associated with the K3 geometry. There are 57 such parameters, $e^m_a$ where $m=1,2,3$ and $a=1,..., 19$. Fluxes fix all $e^m_a$ with $m=1,2$ and $a=1,...,19$. The remaining moduli are $x_i+\bar x_i,\,  i=1, ..., 19$, they originate from $e^3_a$.

In the language of \cite{Aspinwall:2005ad} $x_i+\bar x_i$ are the areas of all $\mathbb{P}^1$ in K3. When the complex structure is stabilized by fluxes, the K3 surface becomes an attractive K3, i.e. the Kummer surface: it has 20 complexified K\"{a}hler forms, 19 $x_i$ and the total volume-axion.

In this model, as different from Calabi-Yau three-folds, we have a product space and therefore the K\"ahler potential is a product of 2 K\"ahler potentials:
\begin{equation}
K=- \ln (T+\bar T)-\ln  [(x_0+\bar x_0)^2-\sum_{i=1}^{i=19}(x_i+\bar x_i)^2] \ .
\end{equation}
Assuming that the K\"ahler volume modulus of the K3 manifold $T+\bar T$ is fixed we can consider the K\"ahler-Hodge manifold $S0(2,19)\over SO(2)\times SO(19)$ with the K\"ahler potential
\begin{equation}
K=\ln \ .
\end{equation}
This K\"ahler potential  is  suitable for our purpose after expanding near the minimum for $x_0+\bar x_0=v$, where, for small $x_i+\bar x_i$, it will look like
\begin{equation}
K\approx {\sum_{i=1}^{i=19}(x_i+\bar x_i)^2\over v^2}\ .
\end{equation}
In previous work \cite{Kallosh:2007ig} we observed that in this situation the exponential terms in the superpotential have to come from the Euclidean instantons and not from the gaugino condensation since the moduli $x_i$ originate from hypermultiplets \cite{Aspinwall:2005ad,Andrianopoli:2003jf}. These exponential terms have the form \cite{Denef:2004ze}
\begin{equation}
e^{-2\pi n x_i} \label{DDF1}
\end{equation}
with integer $n$. Such exponents will not result in a flat axion direction, as we require small factors in the exponent. Thus our conclusion for this example in \cite{Kallosh:2007ig} was not very optimistic.

Now, however, after having explained in section \ref{twopi} the complicated situation with numerical factors in string theory, we must, instead of  (\ref{DDF1}), actually use
\begin{equation}
e^{-{2\pi \over (2 \pi \sqrt{\alpha'})^{4}}\tau} \label{general1}
\end{equation}
since  the units $2\pi \sqrt {\alpha'}=1$ cannot be used, rather one should convert them to $M_{Pl}=1$ units. In fact, taking into account that in \cite{Kachru:2003sx} we had an example with $T_3={2\pi \over g_s( 2 \pi \sqrt{\alpha'})^{4}}\sim 10^{-3} M_{Pl}^2$ one can hope that a small number in the exponent is possible.

Here we also have to make sure that the factors in exponents are not too small so that the multi-instanton correction problem, raised in \cite{Banks:2003sx} and discussed above, should not destroy the model. Here the solution to this problem suggested in section \ref{sugramulti} may work. Indeed we need to have more than one exponent to provide the required periodicity of the axion potential, so that each exponent at the critical point is not small, rather only their difference is. The details of such configurations have to be worked out.

The model with $K3\times {T^2\over \mathbb{Z}_2}$  compactification, therefore, requires additional study before a positive/negative conclusion on the axion valley can be reached. However, we observe that quadratic shift-symmetric  potentials seem to be available, at least in principle, in this particular compactification of string
theory.

\subsubsection{Axions in type IIB orientifolds originating from the 2-form fields}
A closely related program was proposed in \cite{Grimm}, where the possibility of using a general class of type IIB orientifold compactifications \cite{Grimm:2004uq} to construct the axion valley inflationary models was suggested. In these models the involution of the world-sheet parity operation splits the cohomology groups of the compact space into even and odd parts $H^{p,q}_+\oplus H^{p,q}_-$. When $h_-^{1,1}= N$ does not vanish, one can form a set of chiral scalar fields which originate from a combination of the RR and NS-NS 2-form fields, $\hat C_2= c^a \omega_a$ and $\hat B_2= b^a \omega_a$ (with $\{\omega_{a}\}$ a basis for $H^{2}(X,\mathbb{Z})$ and $\tau$ the axion-dilaton)
\begin{equation}
G^a= c^a-\tau b^a \qquad a=1, ..., N \ .
\end{equation}
In  many  models of type IIB orientifold compactification such fields are absent since the 2-forms $\hat C_2$ and $\hat B_2$ are odd under the orientifold symmetry and are thus are projected out. Here, however, since $h_-^{1,1}\neq 0$ the scalars survive and may play an important role in cosmology. The K\"{a}hler potential in the case of one K\"{a}hler moduli $T$, and $N$ $G^a$ fields (with fixed $\tau$ and constants $C_{ab}$) is \cite{Grimm:2004uq}
\begin{equation}
K=-3 \ln[-i(T-\bar T)- C_{ab}(G-\bar G)^a(G-\bar G)^b] \ .
\end{equation}
Such K\"{a}hler potentials may lead to axion-type inflation. Of course, as the first step, one needs show that there is a definite mechanism by which all fields, including $G^{a}$, can be stabilized. Secondly, one should present a concrete example in which a  minimum  can be established. Finally, the possibility of having detectable levels of tensor fluctuations should be investigated in that vacuum via explicit computation. Further, one may also look for models with many $G^a$ fields, i.e. for large $N$ and assisted inflation.

While the details of moduli stabilization still have to be worked out, these models are promising and may eventually lead to viable string inflation with IGW.

\section{Discussion}\label{discussion}
We have explored the possibility of deriving axion inflation (in particular N-flation) in string theory. We do so, noting that such models offer some hope of implementing stringy inflation giving substantial tensor fluctuations and thus measurable levels of CMB polarization. We began in section \ref{nflation} by highlighting some of the problems of embedding inflation (with measurable tensor fluctuations) in string theoretic settings. Following this we examined N-flation in more detail, noting that in most reasonable implementations it is assumed that the volume moduli have significantly steeper potentials than the axions that drive inflation.

In section \ref{onemodulus} we considered the simplest KKLT-type models. We found that, despite the shift symmetry of the K\"{a}hler potential, the curvature of the potential in the volume direction was of the same order as in the axion direction -- i.e. both axion and volume moduli had the same order mass.

In section \ref{kahlerstab} we considered an $\alpha'$ corrected K\"{a}hler potential for many K\"{a}hler moduli. Our motivation here was to have the shift-symmetric $K$ itself stabilize the moduli. By construction, such a scenario could only give a mass to the volume-type moduli, potentially allowing non-perturbative corrections to give small masses to the axions and creating the hierarchy required for N-flation. However, we found several theoretical obstacles; with a constant superpotential, neither SUSY nor (with some caveats) non-SUSY stable vacua were found. Furthermore it appears to be only possible to stabilize one volume modulus using the K\"{a}hler potential alone.  Therefore in this setting it is difficult to justify the   multiple axions rolling in concert to inflate the universe. We also examined some specific examples with a non-constant superpotential, \cite{Conlon:2005ki,Westphal:2006tn}. Although in the large volume example of \cite{Conlon:2005ki} a large mass hierarchy is obtained, it is only for one  of moduli, and even in that case the mass of the axion is too light to provide a realistic model of inflation.

Therefore, at present we find it  difficult within  the standard corner of the stringy landscape\footnote{Under standard corner of the stringy landscape we mean the well studied Calabi-Yau compactifications with unbroken supersymmetry and known structure of the K\"{a}hler potentials. Supersymmetric vacua do not have to come from Calabi-Yau manifolds; some more general examples can be found by considering ``generalized complex'' manifolds.  However, to find  the effective theory around such vacua is difficult \cite{Grana:2005ny}.} to remove the volume moduli from the dynamics and to keep only nearly flat axion directions. More precisely, it seems impossible to  give the axion a substantially lower mass than its
partner. We have pointed out some caveats in our discussions of these issues; these may be studied in future and could, perhaps, change our current conclusions.

Having had no luck with constructible stringy models, we examined supergravity scenarios in section \ref{sugramodels}. Here we use a quadratic  shift-symmetric (axion independent) K\"{a}hler potential and find that  axion pNGb-type inflation with IGW  is possible, as proposed in \cite{Kallosh:2007ig}. Moreover, we have found here a possible solution to the problem of the multi-instanton corrections \cite{Banks:2003sx}: one needs a racetrack-type superpotentials with 2 or more exponents, so that each exponent is not small but the small difference between them may provide the correct parameters for inflationary models with IGW.

Also in section 6 we argued that it may be useful to look for type IIB orientifold models which may have features different from generic Calabi-Yau compactifications: such models provide, in certain approximations, the required quadratic shift-symmetric (axion independent) K\"{a}hler potentials. Often in such orientifold models the axions originate from hypermultiplets.  Whether such models will be valid in the regime  predicting inflationary gravity waves has to be studied and remains a project we postpone for the future.

There are also string models of inflation with IGW which we have not been discussed in this paper\footnote{ For example, a model in \cite{Krause:2007jr} based on heterotic M-theory predicts observable $r$. However, the problem of  stabilization of the orbifold-length and Calabi-Yau volume moduli has not been solved so far. In \cite{Becker:2007ui}, \cite{Kobayashi:2007hm} the DBI-type model \cite{Alishahiha:2004eh}  of a D5 brane wrapped on a 2-cycle in Klebanov-Strassler throat geometry predicts IGW, according to \cite{Becker:2007ui}. However, the problems of back reaction and unusually large orbifolding in  \cite{Becker:2007ui} still have  to be resolved.}.
In all cases, more study will be required.

Given both the results discussed in this paper and other work in this area, we must conclude that finding measurable IGW in stringy models remains problematic. Accordingly, should the next generation of CMB experiments measure appreciable B-mode polarization and thus infer a substantial scalar-tensor ratio, this would pose a significant challenge for string theory.

\section*{Acknowledgments}
We thank S. Kachru and T. Grimm  for  detailed and helpful discussions of various aspects of this work. We are also grateful to T. Banks, B. Florea,  L. Kofman,  A. Linde, J. Maldacena, L. McAllister, F. Quevedo, E. Silverstein,  A. Tomasiello and H. Tye for discussions of the theoretical issues raised in this paper. We thank  R. Blandford, R. Bond, S. Church, G. Efstathiou, B. Netterfield and L. Page who helped us to understand the situation
with current and future observations of B-mode polarization from inflation. This work is supported by the NSF grant 0244728.

\appendix
\section{Type IIB Details}
As outlined in the main body of the text, for the majority of this paper we have been working in the context of IIB string compactifications. Here we flesh out some of the details of our analysis. With complex moduli fluxes stabilized by fluxes and a canonical K\"{a}hler potential, we have:
\begin{eqnarray}\label{appendixpotential}
V=e^{K}\Big(\sum_{i=1}^{n}|D_{i}W|^{2}-3|W|^{2}\Big)\ ,
\end{eqnarray}
where the K\"{a}hler potential is given by
\begin{eqnarray}\label{appendixkahlerpot}
K(\{T_{i},\bar{T}_{i}\})=-2\ln(\hat{\mathcal{V}})=-2\ln\Big((T_{i}+\bar{T}_{i})\hat{v}^{i}\Big)\
,
\end{eqnarray}
where the volume of the two cycles $\hat{v}^{i}$ should be understood as functions of volumes of four cycles, $T_{i}$ and $\bar{T}_{i}$ (the relation between $T_{i}$ and $\hat{v}^{i}$ is given as $\hat{\sigma}_{i}\equiv \mbox{Re}(T_{i})=C_{ijk}\hat{v}^{j}\hat{v}^{k}$, where $C_{ijk}$'s are classical intersection numbers of the Calabi-Yau.). The K\"{a}hler metric and its inverse (derived from the above K\"{a}hler potential) are given by
\begin{eqnarray}\label{met-tem}
&& G^{ij}=-\frac{3}{8\hat{\mathcal{V}}}\Big(M^{ij}-\frac{3\hat{v}^{i}\hat{v}^{j}}{\hat{\mathcal{V}}}\Big)\ ,\\
&& G_{ij}=-\frac{8}{3}\hat{\mathcal{V}}\Big(M_{ij}+\frac{3\hat{\sigma}_{i}
\hat{\sigma}_{j}}{2\hat{\mathcal{V}}}\Big)\ ,
\end{eqnarray}
in which $M_{ij}$ with lower indices is defined as $M_{ij}=C_{ijk}\hat{v}^{k}$ and $M^{ij}$ with upper indices is inverse of $M_{ij}$. The superpotential has the form:
\begin{eqnarray}\label{appendixsuperpot}
W(\{T_{i}\})=W_{0}+W_{1}(\{T_{i}\})\ ,
\end{eqnarray}
with $W_{0}$ the flux superpotential (independent of K\"{a}hler moduli) and $W_{1}$ depends only on the K\"{a}hler moduli. Now, let's calculate the derivatives of the potential and the mass matrix elements:
\begin{eqnarray}\label{DV}
D_{i}V=e^{K}\Big((D_{i}D_{j}W)\bar{D}^{j}\bar{W}-2(D_{i}W)\bar{W}\Big)=0\
.
\end{eqnarray}
We can now easily calculate the holomorphic-holomorphic and holomorphic-antiholomorphic sectors of the mass matrix of the potential:
\begin{eqnarray}
&&D_{i}D_{j}V=e^{K}\Big((D_{i}D_{j}D_{k}W)\bar{D}^{k}\bar{W}-(D_{i}D_{j}W)\bar{W}\Big)\ ,\label{DDV1}\\
&&\bar{D}_{\bar{i}}D_{j}V=e^{K}\Big(-{\mathcal{R}}^{l}_{j\bar{i}k}(D_{l}W)\bar{D}^{k}\bar{W}+G_{\bar{i}j}(D_{k}W)
\bar{D}^{k}\bar{W}-(\bar{D}_{\bar{i}}\bar{W})D_{j}W\nonumber\\
&&\hspace{2.5cm}+(D_{j}D_{k}W)(\bar{D}_{\bar{i}}\bar{D}^{k}\bar{W})-2G_{\bar{i}j}|W|^{2}\Big)\
.\label{DDV2}
\end{eqnarray}
\\\\
\textbf{Stabilization at Supersymmetric Minima}
\\\\
According to the KKLT scenario, the K\"{a}hler moduli are stabilized at supersymmetric minima. The expressions for the mass matrix elements (\ref{DDV1}) and (\ref{DDV2}) are:
\begin{eqnarray}
&&D_{i}D_{j}V=-e^{K}\bar{W}D_{i}D_{j}W\ ,\label{susyDDV1a}\\
&&\bar{D}_{\bar{i}}D_{j}V=e^{K}\Big((D_{j}D_{k}W)(\bar{D}_{\bar{i}}\bar{D}^{k}\bar{W})-2G_{\bar{i}j}|W|^{2}\Big)\
.\label{susyDDV2a}
\end{eqnarray}
Also, it is easy to show that for a supersymmetric minimum we have
\begin{eqnarray}\label{susyDDW}
D_{i}D_{j}W=\partial_{i}\partial_{j}W-\frac{9\hat{v}_{i}\hat{v}_{j}}{{\hat{\mathcal{V}}}^{2}}W+G_{ij}W\ .
\end{eqnarray}
Note that the indices in the second term of (\ref{susyDDW}) are raised with $\delta_{i}^{j}$.

\subsection{A Single K\"{a}hler Modulus}
To understand the basic features of the problem lets restrict ourselves to the simplest case where we only have one K\"{a}hler modulus $n=h^{1,1}(X)=1$. The K\"{a}hler potential is given by
\begin{eqnarray}\label{kklt1kah}
K=-3\ln(T+\bar{T})\ .
\end{eqnarray}
With the above, we can compute the K\"{a}hler connection, K\"{a}hler metric, Levi-Civita connection associated with the K\"{a}hler metric, and the curvature of the K\"{a}hler moduli space. They are given by
\begin{eqnarray}\label{kahgeo}
&&\partial_{T}K=-\frac{3}{T+\bar{T}}\ ,\\
&& G_{\bar{T}T}=\frac{3}{(T+\bar{T})^{2}}\ \ ,\ \ G^{\bar{T}T}=\frac{(T+\bar{T})^{2}}{3}\ ,\\
&&\Gamma^{T}_{TT}=-\frac{2}{T+\bar{T}}\ ,\\
&&
{\mathcal{R}}^{T}_{T\bar{T}T}=\partial_{\bar{T}}\Gamma^{T}_{TT}=\frac{2}{(T+\bar{T})^{2}}\ .
\end{eqnarray}
For this case, (\ref{susyDDW}) reduces to
\begin{eqnarray}\label{susyDDW2}
D_{T}D_{T}W=\partial^{2}_{T}W-\frac{6W}{(T+\bar{T})^{2}}\ ,
\end{eqnarray}
giving
\begin{eqnarray}
&&D_{T}D_{T}V=-\frac{1}{(T+\bar{T})^{3}}\Big(\bar{W}\partial^{2}_{T}W-\frac{6|W|^{2}}{(T+\bar{T})^{2}}\Big)\ ,\label{susyDDV1-1}\\
&&\bar{D}_{\bar{T}}D_{T}V=\frac{1}{(T+\bar{T})^{3}}\Big(\frac{1}{3}(T+\bar{T})^{2}|\partial^{2}_{T}W|^{2}
-2(W\bar{\partial}^{2}_{\bar{T}}\bar{W}+\bar{W}\partial^{2}_{T}W)+\frac{6|W|^{2}}{(T+\bar{T})^{2}}\Big) \ . \label{susyDDV2-2}
\end{eqnarray}
Notice that (\ref{susyDDV2-2}) is purely real, as expected. Now, let's rewrite the K\"{a}hler modulus $T$ in terms of the axion and dilaton fields: $T=\sigma+i\alpha$ and $\bar{T}=\sigma-i\alpha$ where $\sigma$ and $\alpha$ are real fields:
\begin{eqnarray}
&& \frac{\partial^{2}V}{\partial\sigma^{2}}=2D_{T}\bar{D}_{\bar{T}}V+D_{T}D_{T}V+\bar{D}_{\bar{T}}\bar{D}_{\bar{T}}V\ , \label{d2sigma}\\
&& \frac{\partial^{2}V}{\partial\alpha^{2}}=2D_{T}\bar{D}_{\bar{T}}V-(D_{T}D_{T}V+\bar{D}_{\bar{T}}\bar{D}_{\bar{T}}V)\ ,\label{d2alpha}\\
&&
\frac{\partial^{2}V}{\partial\sigma\partial\alpha}=\frac{\partial^{2}V}{\partial\alpha\partial\sigma}=i(D_{T}D_{T}V-
\bar{D}_{\bar{T}}\bar{D}_{\bar{T}}V)\ .\label{d2sigmaalpha}
\end{eqnarray}
Now, using (\ref{susyDDV1-1}) and (\ref{susyDDV2-2}), we find that
\begin{eqnarray}
&&
\frac{\partial^{2}V}{\partial\sigma^{2}}=\frac{24|W|^{2}}{(T+\bar{T})^{5}}-
\frac{5(W\bar{\partial}^{2}_{\bar{T}}\bar{W}+\bar{W}{\partial}^{2}_{T}W)}{(T+\bar{T})^{3}}+
\frac{2|\partial^{2}_{T}W|^{2}}{3(T+\bar{T})}\ ,\label{d2sigma-2}\\
&&
\frac{\partial^{2}V}{\partial\alpha^{2}}=\frac{2|\partial^{2}_{T}W|^{2}}{3(T+\bar{T})}-
\frac{3(W\bar{\partial}^{2}_{\bar{T}}\bar{W}+\bar{W}{\partial}^{2}_{T}W)}{(T+\bar{T})^{3}}\ ,\label{d2alpha-2}\\
&&
\frac{\partial^{2}V}{\partial\sigma\partial\alpha}=i\frac{W\bar{\partial}^{2}_{\bar{T}}\bar{W}-
\bar{W}\partial^{2}_{T}W}{(T+\bar{T})^{3}}\ .\label{d2sigmaalpha-2}
\end{eqnarray}
The masses of the $\sigma$ and $\alpha$ fields will then correspond to the eigenvalues of the following 2-dimensional symmetric matrix:
\begin{eqnarray}\label{d2Va}
H=\left(
    \begin{array}{cc}
      \partial_{\sigma}\partial_{\sigma}V & \partial_{\sigma}\partial_{\alpha}V \\
      \partial_{\alpha}\partial_{\sigma}V & \partial_{\alpha}\partial_{\alpha}V \\
    \end{array}
  \right)\ .
\end{eqnarray}
These eigenvalues are:
\begin{eqnarray}
&&
\lambda_{1}=\frac{H_{11}+H_{22}+\sqrt{(H_{11}-H_{22})^{2}+4H_{12}^{2}}}
{2}\ ,\label{eigenvalue1}\\
&&
\lambda_{2}=\frac{H_{11}+H_{22}-\sqrt{(H_{11}-H_{22})^{2}+4H_{12}^{2}}}
{2}\ .\label{eigenvalue2}
\end{eqnarray}

\subsubsection{KKLT}
In the KKLT model, the superpotential is given by
\begin{eqnarray}\label{kkltsuppot1}
W(T)=W_{0}+Ae^{-aT}\ ,
\end{eqnarray}
where $A$ and $a$ are some constants (we also assume that they are real). Since we are interested in evaluating (\ref{susyDDV1-1}) and (\ref{susyDDV2-2}) at the supersymmetric critical point, we can simplify these equations. First, we notice that at such a point we have
\begin{eqnarray}\label{eaT}
e^{-aT}=-\frac{3W_{0}}{A(3+a(T+\bar{T}))}\ ,\ \ \
e^{-a\bar{T}}=-\frac{3\bar{W}_{0}}{A(3+a(T+\bar{T}))}\ .
\end{eqnarray}
If we combine these two minimization equations, we can find the values of $\sigma$ and $\alpha$ fields at the stabilization point as
\begin{eqnarray}\label{sigalp1}
e^{-a(T+\bar{T})}=\frac{9|W_{0}|^{2}}{A^{2}(3+a(T+\bar{T}))^{2}}\ ,\
\ e^{-a(T-\bar{T})}=\frac{W_{0}}{\bar{W}_{0}}\ ,
\end{eqnarray}
which gives the stabilization value of $\sigma$ field as
\begin{eqnarray}\label{sigalp2a}
e^{-2a\sigma_{0}}=\frac{9|W_{0}|^{2}}{A^{2}(3+2a\sigma_{0})^{2}}\
.
\end{eqnarray}
Note that the above equation cannot be solved analytically. For the $\alpha$ field, we obtain
\begin{eqnarray}\label{sigalp3a}
\alpha_{0}=-\frac{\theta}{a}+\frac{n\pi}{a}\ ,\
n\in{\mathbb{Z}}\ ,
\end{eqnarray}
where $\theta$ is defined as the phase of the flux superpotential $e^{i\theta}=\frac{W_{0}}{|W_{0}|}$. The interesting fact about (\ref{sigalp3a}) is that the value of $\alpha_{0}$ is fixed up to a shift symmetry. Now, let's calculate the matrix elements at the stabilization point. It is easy to show that
\begin{eqnarray}\label{DDV-DDV1}
D_{T}D_{T}V-\bar{D}_{\bar{T}}\bar{D}_{\bar{T}}V=a^{2}A\frac{W_{0}e^{-a\bar{T}}-\bar{W}_{0}e^{-aT}}{(T+\bar{T})^{3}}\
.
\end{eqnarray}
At the supersymmetric minima this vanishes, from (\ref{eaT}):
\begin{eqnarray}\label{DDV-DDV-V1}
D_{T}D_{T}V-\bar{D}_{\bar{T}}\bar{D}_{\bar{T}}V=0\ .
\end{eqnarray}
This implies that the off-diagonal elements of the $H$ matrix vanish ($H_{12}=H_{21}=0$). Remembering (\ref{eigenvalue1}) and (\ref{eigenvalue2}), the masses of the two fields at the supersymmetric minima are given by
\begin{eqnarray}\label{mass1}
m_{\sigma}^{2}=\sigma_{0}^{2}H_{11}\ , \ \
m_{\alpha}^{2}=\sigma_{0}^{2}H_{22}\ .
\end{eqnarray}
Note in above that in order to get the actual mass difference with the correct normalized kinetic terms, we have multiplied the two eigenvalues $\lambda_{1}$ and $\lambda_{2}$ by a factor of $\sigma_{0}^{2}$. Now, if we substitute the KKLT superpotential (\ref{kkltsuppot}) for (\ref{d2sigma-2}) and (\ref{d2alpha-2}) and use the fact that at the supersymmetric minima we have (\ref{eaT}) and (\ref{sigalp1}), then the masses of the two fields are
\begin{eqnarray}
&&
m_{\sigma}^{2}=\frac{3|W_{0}|^{2}}{8\sigma_{0}^{3}}\frac{4(a\sigma_{0})^{2}(2+5(a\sigma_{0})+2(a\sigma_{0})^{2})}
{(3+2a\sigma_{0})^{2}}\ ,\label{masssigma1}\\
&&
m_{\alpha}^{2}=\frac{3|W_{0}|^{2}}{8\sigma_{0}^{3}}\frac{4(a\sigma_{0})^{3}}{3+2a\sigma_{0}}\
.\label{massalpha1}
\end{eqnarray}
We notice that all masses are independent of the value of $\alpha_{0}$ and only depend on the value of $\sigma_{0}$ and $|W_{0}|$. It should also be noticed that $|W_{0}|$ appears in both masses in the same way. The ratio of the masses is given by
\begin{eqnarray}\label{ratio1}
\frac{m_{\sigma}^{2}}{m_{\alpha}^{2}}=1+\frac{2(1+a\sigma_{0})}{(a\sigma_{0})(3+2a\sigma_{0})}\
.
\end{eqnarray}

\subsubsection{KL}
The superpotential of this model is given by:
\begin{eqnarray}\label{klsuppot1}
W(T)=W_{0}+Ae^{-aT}+Be^{-bT}\ ,
\end{eqnarray}
where $a$, $b$, $A$, and $B$ are real constants. The minimization conditions for the supersymmetric vacua ($D_{T}W=0$ and $\bar{D}_{\bar{T}}\bar{W}=0$) are
\begin{eqnarray}
&&A(3+a(T+\bar{T}))e^{-aT}+B(3+b(T+\bar{T}))e^{-bT}=-3W_{0}\ ,\label{stab1}\\
&&A(3+a(T+\bar{T}))e^{-a\bar{T}}+B(3+b(T+\bar{T}))e^{-b\bar{T}}=-3\bar{W}_{0}\
.\label{stab2}
\end{eqnarray}
Now, using (\ref{d2sigma-2}), (\ref{d2alpha-2}), and (\ref{d2sigmaalpha-2}), we obtain for the mass matrix elements the following
\begin{eqnarray}
&&
\sigma_{0}^{2}H_{11}=\frac{1}{6\sigma_{0}^{3}}\Bigg\{(a\sigma_{0})^{2}A^{2}e^{-2a\sigma_{0}}(2+a\sigma_{0})
(1+2a\sigma_{0})+(b\sigma_{0})^{2}B^{2}e^{-2b\sigma_{0}}(2+b\sigma_{0})
(1+2b\sigma_{0})\nonumber\\
&&\hspace{1.6cm}
+(a\sigma_{0})(b\sigma_{0})ABe^{-(a+b)\sigma_{0}}\big(4+5(a\sigma_{0}+b\sigma_{0})+4(a\sigma_{0})(b\sigma_{0})\big)
\cos((a-b)\alpha_{0})\Bigg\}\ ,\label{klh11}\\
&&
\sigma_{0}^{2}H_{22}=\frac{1}{6\sigma_{0}^{3}}\Bigg\{(a\sigma_{0})^{3}A^{2}e^{-2a\sigma_{0}}(3+2a\sigma_{0})
+(b\sigma_{0})^{3}B^{2}e^{-2b\sigma_{0}}(3+2b\sigma_{0})\nonumber\\
&&\hspace{1.6cm}+(a\sigma_{0})(b\sigma_{0})ABe^{-(a+b)\sigma_{0}}\big(3(a\sigma_{0}+b\sigma_{0})+4(a\sigma_{0})
(b\sigma_{0})\big)\cos((a-b)\alpha_{0})\Bigg\}\ ,\label{klh22}\\
&&
\sigma_{0}^{2}H_{12}=\frac{1}{\sigma_{0}^{3}}\Big((a\sigma_{0})(b\sigma_{0})(a\sigma_{0}-b\sigma_{0})AB
e^{-(a+b)\sigma_{0}}\sin((a-b)\alpha_{0})\Big)\ ,\label{klh12}
\end{eqnarray}
which are still subject to the constraints (\ref{stab1}) and (\ref{stab2}). There are several important facts about these expressions. First, note that we have an off-diagonal term and the masses of $\sigma$ and $\alpha$ fields will correspond to the eigenvalues of $H$. The second thing is that if we put $A=0$ (or $B=0$) or $a=b$, then these expressions clearly reduce to KKLT answers. It is straightforward, if unilluminating, to diagonalize the mass matrix and find the dilaton/axion mass ratio. A more fruitful way to proceed is consider the relevant limits for our problem, as discussed in the main body of the text.

\section{Derivation of the $\alpha'$ Corrected Potential}
In this section, we explicitly calculate the ${\mathcal{N}}=1$ supergravity potential of type IIB string theory compactifications on a Calabi-Yau threefold $Z$ in the presence of $\alpha'$ corrections. The potential of the theory is
\begin{eqnarray}\label{z-12}
V=e^{K}\Big((G^{-1})^{I\bar{J}}D_{I}W
D_{\bar{J}}\bar{W}-3|W|^{2}\Big)\ ,
\end{eqnarray}
where index $I$ runs over all complex structure and K\"{a}hler moduli as well as the axion-dilaton field. The superpotential $W$ is the flux superpotential and is only a function of complex structure moduli and the axion-dilaton. The superpotential does not receive any $\alpha'$ corrections. The K\"{a}hler potential, however, does receive corrections and it is given by
\begin{eqnarray}\label{z-5}
K=\ln(-i(\tau-\bar{\tau}))-2\ln\Bigg[2\hat{\sigma}_{i}\hat{v}^{i}+\xi\Big(
-\frac{i(\tau-\bar{\tau})}{2}\Big)^{3/2}\Bigg]-\ln\Bigg[-i\int_{Z}\Omega\wedge\bar{\Omega}\Bigg]\
,
\end{eqnarray}
where $\hat{v}^{i}=e^{-\phi_{0}/2}v^{i}$ and $\hat{\sigma}_{i}=e^{-\phi_{0}}\sigma_{i}$ are the volumes of the two and four cycles of the Calabi-Yau respectively. We notice that the K\"{a}hler moduli are defined in terms of the volumes of the four cycles $-i(T_{i}-\bar{T}_{i})=2\hat{\sigma}_{i}$, and that the $\hat{v}^{i}$ in (\ref{z-5}) should be understood as a function of $\hat{\sigma}_{i}$'s. Now, in order to calculate the potential, we should first calculate the Weil-Peterson metric derived from the above K\"{a}hler potential.

From (\ref{z-5}), it is clear that there are off diagonal components of the moduli space metric which mix the K\"{a}hler moduli with the axion-dilaton. Before calculating the metric and its inverse, let us consider a more general problem and then apply it for our specific problem. Consider the following symmetric $(n+1)\times(n+1)$ dimensional invertible matrix $Q$
\begin{eqnarray}\label{z-1}
Q=\left(
    \begin{array}{cc}
      A_{1\times1} & B_{1\times n} \\
      B^{\mathsf{T}}_{n\times1} & Y_{n\times n} \\
    \end{array}
  \right)\ ,
\end{eqnarray}
where $Y$ is a symmetric $n\times n$ matrix. Assume that the inverse of $Y$ is given. Then we want to find the inverse of $Q$ in terms of $Y^{-1}$. Without loss of generality, we can assume that $Q^{-1}$ has the following form
\begin{eqnarray}\label{z-2}
Q^{-1}=\left(
         \begin{array}{cc}
           A^{'} & B^{'} \\
           B^{'\mathsf{T}} & Y^{-1}+D \\
         \end{array}
       \right)\ ,
\end{eqnarray}
where $A'$, $B'$ and $D$ should be determined in terms of $A$, $B$, and $Y^{-1}$. Requiring that $Q^{-1}$ is the inverse of $Q$; i.e. $QQ^{-1}=Q^{-1}Q=1_{(n+1)\times(n+1)}$, we obtain the following equations
\begin{eqnarray}
&& AA^{'}+BB^{'\mathsf{T}}=1\ ,\label{z-3-1}\\
&& AB^{'}+BY^{-1}+BD=0\ ,\label{z-3-2}\\
&& B^{\mathsf{T}}A^{'}+YB^{'\mathsf{T}}=0\ ,\label{z-3-3}\\
&& B^{\mathsf{T}}B^{'}+YD=0\ .\label{z-3-4}
\end{eqnarray}

From (\ref{z-3-4}), we immediately find that $D=-Y^{-1}B^{\mathsf{T}}B^{'}$. Substituting this into (\ref{z-3-2}), we find $B^{'}=\frac{BY^{-1}}{BY^{-1}B^{\mathsf{T}}-A}$. Applying this result for (\ref{z-3-3}), we find $A'$ as $A^{'}=\frac{-1}{BY^{-1}B^{\mathsf{T}}-A}$. Equation (\ref{z-3-1}) is trivially satisfied. Therefore, we find the inverse of $Q$ as the following
\begin{eqnarray}\label{z-4}
Q^{-1}=\frac{1}{BY^{-1}B^{\mathsf{T}}-A} \left(
\begin{array}{cc}
-1 & BY^{-1} \\
Y^{-1}B^{\mathsf{T}} & (BY^{-1}B^{\mathsf{T}})Y^{-1}-(Y^{-1}B^{\mathsf{T}})(BY^{-1})-AY^{-1} \\
\end{array}
\right)\ .
\end{eqnarray}

Now, we are ready to apply this result to our main problem. Before doing so, we introduce some notation to simplify the calculations. We define $M_{ij}=C_{ijk}\hat{v}^{k}$ and $M=\hat{\mathcal{V}}+\frac{1}{2}\hat{\xi}$ where $\hat{{\mathcal{V}}}=\hat{\sigma}_{i}\hat{v}^{i}$. Recalling
$\hat{\sigma}_{i}=C_{ijk}\hat{v}^{j}\hat{v}^{k}=M_{ij}\hat{v}^{j}$ and $\frac{\partial}{\partial T_{i}}=-\frac{i}{2}\frac{\partial}{\partial\hat{\sigma}_{i}}$, we obtain
\begin{eqnarray}\label{z-6}
\frac{\partial\hat{v}^{i}}{\partial\hat{\sigma}_{j}}=\frac{1}{2}M^{ij}\
,
\end{eqnarray}
where $M^{ij}=(M^{-1})^{ij}$ with upper indices should be understood as the inverse of $M_{ij}$. Now, we identify $Y$ in (\ref{z-1}) by $Y^{ij}=\frac{\partial^{2}K}{\partial T_{i}\partial \bar{T}_{j}}$. After some calculation, we find
\begin{eqnarray}\label{z-7}
Y^{ij}=-\frac{3}{8M}\Big(M^{ij}-\frac{3\hat{v}^{i}\hat{v}^{j}}{M}\Big)\
.
\end{eqnarray}
It is easy to check that $(Y)^{-1}$ (which we display with lower indices) is given by
\begin{eqnarray}\label{z-8}
Y_{ij}=-\frac{8}{3}M\Big(M_{ij}-\frac{6\hat{\sigma}_{i}\hat{\sigma_{j}}}{4M-3\hat{\xi}}\Big)\
.
\end{eqnarray}

The last step before using the above prescription (\ref{z-4}), is to identify $A$ and $B$. They are given by the following expressions
\begin{eqnarray}
&& A=\partial_{\tau}\partial_{\bar{\tau}}K=\frac{1}{4}e^{2\phi_{0}}\Big(1-\frac{3\hat{\xi}}{8M^{2}}(2M-3\hat{\xi})\Big)\ ,\label{z-9-1}\\
&&
B^{i}=\partial_{\bar{\tau}}\partial_{T_{i}}K=\frac{9}{16}e^{\phi_{0}}\hat{\xi}\frac{\hat{v}^{i}}{M^{2}}\
.\label{z-9-2}
\end{eqnarray}
We have completely constructed the metric of the K\"{a}hler moduli space and the axion-dilaton
\begin{eqnarray}\label{z-10}
G^{A\bar{B}}=\left(
               \begin{array}{cc}
                 A & B^{i} \\
                 B^{\mathsf{T}i} & Y^{ij} \\
               \end{array}
             \right)\ ,
\end{eqnarray}
and want to obtain its inverse using (\ref{z-4}). Rewriting everything in terms of the volume $\hat{\mathcal{V}}$ of the Calabi-Yau $Z$, we find the various components of the inverse metric as
\begin{eqnarray}
&& (G^{-1})_{\tau\bar{\tau}}=e^{-2\phi_{0}}\frac{4\hat{\mathcal{V}}-\hat{\xi}}{\hat{\mathcal{V}}-\hat{\xi}}\ ,\label{z-11-1}\\
&& (G^{-1})_{\tau\bar{i}}=-3ie^{-\phi_{0}}\frac{\hat{\xi}}{\hat{\mathcal{V}}-\hat{\xi}}\hat{\sigma}_{i}\ ,\label{z-11-2}\\
&&
(G^{-1})_{i\bar{j}}=-\frac{4}{3}(2\hat{\mathcal{V}}+\hat{\xi})M_{ij}+\frac{4\hat{\mathcal{V}}-\hat{\xi}}
{\hat{\mathcal{V}}-\hat{\xi}}\hat{\sigma}_{i}\hat{\sigma}_{j}\
,\label{z-11-3}
\end{eqnarray}
which is in agreement with \cite{Bobkov:2004cy}. Using this inverse metric and plugging into (\ref{z-12}) and using the fact that the superpotential is independent of the K\"{a}hler moduli, the potential then reads
\begin{eqnarray}\label{z-13}
V(\hat{\sigma}_{i})&=&e^{K}\Bigg[(G^{-1})^{\alpha\bar{\beta}}D_{\alpha}W
D_{\bar{\beta}}\bar{W}+e^{-2\phi_{0}}\frac{4\hat{\mathcal{V}}-\hat{\xi}}{\hat{\mathcal{V}}-\hat{\xi}}D_{\tau}W
D_{\bar{\tau}}\bar{W}\nonumber\\
&&\hspace{0.4cm}+\frac{9e^{-\phi_{0}}\hat{\xi}\hat{\mathcal{V}}}{(\hat{\mathcal{V}}-\hat{\xi})
(2\hat{\mathcal{V}}+\hat{\xi})}(WD_{\bar{\tau}}\bar{W}+\bar{W}D_{\tau}W)+3\hat{\xi}
\frac{\hat{\mathcal{V}}^{2}+7\hat{\xi}\hat{\mathcal{V}}+\hat{\xi}^{2}}{(\hat{\mathcal{V}}-\hat{\xi})
(2\hat{\mathcal{V}}+\hat{\xi})^{2}}|W|^{2}\Bigg]\ .
\end{eqnarray}
This was first found in \cite{Becker:2002nn}.

\section{K\"{a}hler Stabilization of the Volume Moduli}
In this section, we present the details of the computations of the K\"{a}hler stabilization mechanism. As mentioned earlier the K\"{a}hler moduli can be stabilized in both supersymmetric and non-supersymmetric vacua. Here, we consider both possibilities.

\subsection{Supersymmetric minima of the potentials}
Here we demonstrate that supersymmetric minima are only possible at vanishing volumes of the 4-cycles. First we notice that the dependence of both K\"{a}hler potentials (\ref{K}) and (\ref{K2}) on K\"{a}hler moduli is only through the volume ${\cal{V}}$. On the other hand, the supersymmetric
critical points of both potentials, for a constant superpotential, correspond to $D_{i}W=(\partial_{i}K)W_{0}=0$. Therefore, the supersymmetric critical points are given by $\partial_{i}K_{1}({\cal{V}})=0$ and $\partial_{i}K_{2}({\cal{V}})=0$. Obviously, both equations reduce to solving $\partial_{i}{\cal{V}}=0$. This is not an acceptable point of the moduli space since we are looking for stabilization with some at least modestly large values of volume moduli where the supergravity approximation is valid. Here we stress that the form of the prepotential in (\ref{large}), where the world-sheet instanton corrections are negligible, required a ``large moduli approximation''. In the deep interior of Calabi-Yau moduli space the
expansion given in (\ref{large}) is not valid.

We conclude that there is no consistent supersymmetric stabilization of the volume moduli starting with (\ref{K}) in the regime of the validity of this equation without introducing the moduli dependence into the superpotential. Analogous considerations are valid in the case of the K\"{a}hler potential in (\ref{K2}), and we reach the same conclusion.

\subsection{Non-supersymmetric minima of the potentials}
Having illustrated non-existence of SUSY minima, we now consider the non-SUSY case. We will consider (\ref{K}) and (\ref{K2}) separately, beginning with the latter. In both cases we find that non-supersymmetric stabilization is only possible for one combination of the moduli, namely ${\cal V}$.

For the type IIB flux compactification models described in (\ref{K1}) the potential has been calculated in \cite{Becker:2002nn}. At the point where the axion-dilaton and the complex structure moduli are fixed by $D_\tau W= D_z W=0$ it is given by
\begin{eqnarray}\label{potential1}
V=3\hat{\xi}
\frac{\hat{\mathcal{V}}^{2}+7\hat{\xi}\hat{\mathcal{V}}+\hat{\xi}^{2}}{(\hat{\mathcal{V}}-\hat{\xi})
(2\hat{\mathcal{V}}+\hat{\xi})^{2}}|W|^{2}\ .
\end{eqnarray}
The complete derivation of this potential is presented above. Here we see that the potential for vanishing $\xi$ is scale-invariant and is identically zero when the superpotential is independent of the volume moduli. This is despite an apparent contradiction due to the presence of the factor $-2$ from the $\ln$ in ((\ref{K1}). In fact, although $-2 \ln {\cal V}$ looks like a square of the cubic in $v$-fields expression, one has to use the dual variables $\tau_i= {1\over 2}C_{ijk}v^j v^k$ which makes the K\"{a}hler potential effectively cubic in $\tau_i$-fields and thus
gives a no-scale potential for vanishing $\xi$.

As we can see, the potential depends on the volume moduli $t^i-\bar t^i$ only via ${\cal V}$. In fact, this can be easily explained from the very special geometry point of view on $N=2$ supergravities in 5d where the theory is defined by the surface \cite{Gunaydin:1983bi}
\begin{equation}
{\cal V}= { 1 \over 3!}c_{ijk} v^i v^j v^k=C\ . \label{surface}
\end{equation}
$v^i$ are the real coordinates of the very special geometry and $c_{ijk}$ (a constant, totally symmetric tensor) defines the Chern-Simons term of the five-dimensional supergravity. ${\cal V}=C$ is an invariant surface defined by the equation (\ref{surface}) and the potential depends on the moduli only via ${\cal V}$.

We are looking at the minima of the potential trying to stabilize all moduli $t^i-\bar t^i$. The first derivative of the potential
(\ref{potential1}) is
\begin{eqnarray}\label{a-dV}
\partial_{i}V=-\frac{18e^{K_{0}}\hat{\xi}}{\kappa^{2}}|W_{0}|^{2}(\partial_{i}\hat{{\mathcal{V}}})\hat{{\mathcal{V}}}
\frac{2\hat{{\mathcal{V}}}^{2}+17\hat{\xi}\hat{{\mathcal{V}}}-10\hat{\xi}^{2}}{(\hat{{\mathcal{V}}}-\hat{\xi})^{2}
(2\hat{{\mathcal{V}}}+\hat{\xi})^{5}}
\end{eqnarray}

$\partial_{i}\hat{{\mathcal{V}}}=0$ is not an acceptable solution, since it would give $\hat{{\mathcal{V}}}=0$ -- we would be far away from the region of moduli space where our calculations are valid. The other, non-supersymmetric, solution requires that
\begin{equation}
{2\hat{{\mathcal{V}}}^{2}+17\hat{\xi}\hat{{\mathcal{V}}}-10\hat{\xi}^{2}}=0\
.
\end{equation}
One can solve this equation and find two solutions
\begin{eqnarray}\label{non-susysol}
\hat{{\mathcal{V}}}_{0}=-\frac{\hat{\xi}}{4}(17+3\sqrt{41})=-9.052\
\hat{\xi}\ ,\ \hat{{\mathcal{V}}}_{0}=
\frac{\hat{\xi}}{4}(-17+3\sqrt{41})=0.552\ \hat{\xi}\ .
\end{eqnarray}
We need to investigate the stability of the above critical points. The second derivatives of the potential at the critical points are given by
\begin{eqnarray}\label{a-mass}
\partial^{2}_{\sigma}V\Big|_{\sigma=\sigma_{0}}=0.001\frac{e^{K_{0}}|W_{0}|^{2}}{\kappa^{2}}\frac{1}{\hat{\xi}^{8/3}}\
,\
\partial^{2}_{\sigma}V\Big|_{\sigma=\sigma_{0}}=-28.46\frac{e^{K_{0}}|W_{0}|^{2}}{\kappa^{2}}\frac{1}{\hat{\xi}^{8/3}}\
.
\end{eqnarray}
As is clear, regardless of sign of $\hat{\xi}$, the first critical point is always a minimum and the second one is a maximum. But since we require to have a positive volume then this implies that $\hat{\xi}$ should be negative for the first critical point. Nevertheless, the first critical point does not have a desirable critical value -- the volume is stabilized at the string scale and again we are in trouble. Moreover, all but one moduli remain flat directions of the potential. The former of these points means that K\"{a}hler stabilization of moduli cannot be performed for the values of the moduli outside the deep interior of the moduli space and the latter means that, at most, only one modulus could be stabilized anyway.

Now let us consider (\ref{K}). Here the potential derived from the above K\"{a}hler potential given by
\begin{eqnarray}\label{Vmod1}
V=3\xi\frac{|W_{0}|^{2}}{({\mathcal{V}}-\xi)({\mathcal{V}}+\xi/2)}\
.
\end{eqnarray}
The only nontrivial critical point of the above potential is the non-supersymmetric critical point ($\partial_{t}V=0$). For this critical point, it turns out that the stabilized volume is given by
\begin{eqnarray}\label{stabV}
{\mathcal{V}}=\frac{\xi}{4}\ ,
\end{eqnarray}
with the second derivative at the critical point $\partial^{2}_{t}V=3.6\ \xi^{-7/3}$. As we see, this critical point is a minimum when $\xi<0$ and is a maximum otherwise. Once again we fail to find an phenomenologically viable axion valley -- a minima would require a negative volume, which is clearly forbidden. Further, as above, the best K\"{a}hler stabilization can do is fix a single volume moduli -- this is insufficient for N-flating the
universe.

Thus, for both (\ref{K}) and (\ref{K2}) we cannot find a suitable starting point for building an axion valley model. In both cases K\"{a}hler stabilization gives a suitable minima; however, said minima are located at points in moduli space that cannot be realized within the regions where the constructions of the K\"{a}hler potentials are valid. Not only that, but only one moduli is stabilized and so a volume/axion hierarchy can only be
established for one moduli. This is not enough to realize N-flation.

\subsubsection{Stabilization at $D_{\tau}W\neq0,\ D_{\alpha}W\neq0$}
In the previous section, we stabilized the axion-dilaton and complex structure moduli at critical points where $D_{\tau}W=0$ and $D_{\alpha}W=0$. This leads to a simplified potential for the K\"{a}hler moduli. We found that this potential has two critical points and one of them is a local minimum. However, the volume of the Calabi-Yau is stabilized at string scale, whereas we are interested only in large volume compactification. The goal of what
follows is to investigate whether such a large volume stabilization can be achieved by relaxing the stabilization conditions $D_{\tau}W=0$ and $D_{\alpha}W=0$.

In this approach \cite{Denef:2004ze}, we ignore the K\"{a}hler moduli at first and take the following potential
\begin{eqnarray}\label{DDpot}
V_{1}=e^{K_{1}}\Big(\sum_{A\in\{\tau,z_{\alpha}\}}|D_{A}W|^{2}-3|W|^{2}\Big)\
,
\end{eqnarray}
where $K_{1}$ is the K\"{a}hler potential of the complex structure moduli space and the axion-dilaton. We then stabilize the axion-dilaton and the complex structure moduli at non-supersymmetric critical points of $V_{1}$. The stabilization equations are non-supersymmetric solutions of $D_{A}V_{1}=0$\ \footnote{\ The explicit form of the equations are given in \cite{Denef:2004ze} and \cite{Soroush:2007ed}.}. However, there is an important fact here about the stability of these solutions. One needs to check explicitly whether these solutions are perturbatively stable. The
complete analysis of the stability of almost supersymmetric vacua (supersymmetry is broken at low scales of energy) can be found in \cite{Denef:2004ze}. However, finding meta-stable non-supersymmetric solutions is generically hard and there are stringent constraints in certain limits. For example, it was explicitly shown in \cite{Soroush:2007ed} that in the dilaton domination limit, there is no meta-stable de Sitter flux vacuum and for anti de Sitter solutions, the stability requirement imposes certain relations between the value of cosmological constant and the scale of supersymmetry breaking. There are also constraints in other regimes. For instance, it was shown in \cite{Gomez-Reino:2006dk} that the curvature of the complex structure moduli space needs to satisfy certain constraints to obtain meta-stable non-supersymmetric vacua.

Here, we do not give any explicit examples of those meta-stable non-supersymmetric vacua which are hard to find. We only investigate whether, if such vacua exist, it is possible to achieve large volume stabilization. The potential for the K\"{a}hler moduli with a stabilized axion-dilaton and complex structure moduli at metastable non-supersymmetric vacuum is given by
\begin{eqnarray}\label{vnonsusy}
V&=&C\Bigg[N_{1}\frac{1}{(2\hat{\mathcal{V}}+\hat{\xi})^{2}}+N_{2}\frac{4\hat{\mathcal{V}}-\hat{\xi}}{(\hat{\mathcal{V}}-
\hat{\xi})(2\hat{\mathcal{V}}+\hat{\xi})^{2}}+N_{3}\frac{\hat{\xi}\hat{\mathcal{V}}}{(\hat{\mathcal{V}}-\hat{\xi})
(2\hat{\mathcal{V}}+\hat{\xi})^{3}}\nonumber\\
&&\hspace{0.6cm}+3|W_{0}|^{2}\hat{\xi}\frac{\hat{\mathcal{V}}^{2}+7\hat{\xi}\hat{\mathcal{V}}
+\hat{\xi}^{2}}{(\hat{\mathcal{V}}-\hat{\xi})(2\hat{\mathcal{V}}+\hat{\xi})^{4}}\Bigg]\
,
\end{eqnarray}
where $C$ is some positive constant and $N_{1}$, $N_{2}$, and $N_{3}$ are given by
\begin{eqnarray}\label{constants}
&& N_{1}=(\bar{D}^{\alpha}\bar{W})_{0}(D_{\alpha}W)_{0}\ ,\\
&& N_{2}=e^{-2\phi_{0}}|(D_{\tau}W)_{0}|^{2}\ ,\\
&&
N_{3}=9e^{-\phi_{0}}(\bar{W}_{0}(D_{\tau}W)_{0}+W_{0}(D_{\bar{\tau}}\bar{W})_{0})\
.
\end{eqnarray}
Here a subscript 0 indicates the value at the non-supersymmetric stabilization point. Now, we should minimize $V$ in terms of K\"{a}hler moduli ($\partial_{i} V=0$). As before, $\partial_{i}\hat{\mathcal{V}}=0$ is trivial and the only nontrivial solution is an algebraic equation in terms of $\hat{\mathcal{V}}$. In this case, we get a quartic equation which is tedious. However, we can simplify the equation in the limit we are interested, namely $\hat{\mathcal{V}}\gg{\hat{\xi}}$. In this limit, the volume gets stabilized at
\begin{eqnarray}\label{nuhat}
\hat{\mathcal{V}}_{0}=\hat{\xi}\frac{4N_{1}-2N_{2}-6N_{3}-9|W_{0}|^{2}}{4(N_{1}+3N_{2})}=\hat{\xi}
\Big(1-\frac{14N_{2}+6N_{3}+9|W_{0}|^{2}}{4(N_{1}+3N_{2})}\Big)\ .
\end{eqnarray}

First notice that dependence on $g_{s}$ cannot help us obtain large volumes, because if we demand small $g_{s}=e^{-\phi_{0}}$, then the only dominant factor will be $N_{2}$ which appears both in numerator and denominator of (\ref{nuhat}). Since $N_{1}$ and $N_{2}$ are positive definite, a large volume is achieved only when \{$\hat{\xi}<0$, $N_{3}\gg N_{1}$, $N_{3}\gg N_{2}$\} or \{$\hat{\xi}<0$, $|W_{0}|^{2}\gg N_{1},N_{2}$\}. The later case implies that we should necessarily be in an anti de Sitter vacuum and therefore is uninteresting. However, the former case describes a de Sitter vacuum. Requiring $N_{3}\gg N_{1}$, it implies that $|D_{\tau}W|\gg|D_{\alpha}W|$, namely the dilaton dominated limit. On the other hand, it was shown in \cite{Soroush:2007ed} that there is no meta-stable de Sitter vacuum in this limit and all vacua are perturbatively unstable. Therefore, It is not possible to construct a K\"{a}hler stabilized model via the K\"{a}hler stabilization procedure in the large volume limit.


\begin{thebibliography}{999}
\bibitem{Efstathiou:2006ak}
  G.~Efstathiou and S.~Chongchitnan,
 ``The search for primordial tensor modes,''
  Prog.\ Theor.\ Phys.\ Suppl.\  {\bf 163}, 204 (2006)
  [arXiv:astro-ph/0603118].

\bibitem{Tegmark:2006az}
  M.~Tegmark {\it et al.},
  ``Cosmological Constraints from the SDSS Luminous Red Galaxies,''
  Phys.\ Rev.\  D {\bf 74}, 123507 (2006)
  [arXiv:astro-ph/0608632].

\bibitem{Pogosian:2006hg}
  L.~Pogosian, I.~Wasserman and M.~Wyman,
``On vector mode contribution to CMB temperature and polarization from  local
  strings,''
  arXiv:astro-ph/0604141.

  \bibitem{Netterfield} B. Netterfield, talk at COSMO 06, http://cosmo06.ucdavis.edu/dltalks.html and private communication

\bibitem{Bock:2006yf}
  J.~Bock {\it et al.},
  ``Task Force on Cosmic Microwave Background Research,''
  arXiv:astro-ph/0604101.



\bibitem{Linde:1983gd}
  A.~D.~Linde,
  ``Chaotic Inflation,''
  Phys.\ Lett.\  B {\bf 129}, 177 (1983).


\bibitem{Destri:2007pv}
  C.~Destri, H.~J.~de Vega and N.~G.~Sanchez,
``MCMC analysis of WMAP3 data points to broken symmetry inflaton potentials
and provides a lower bound on the tensor to scalar ratio,''
  arXiv:astro-ph/0703417.
\bibitem{Kallosh:2007wm}
  R.~Kallosh and A.~Linde,
  ``Testing String Theory with CMB,''
  JCAP {\bf 0704}, 017 (2007)
  [arXiv:0704.0647 [hep-th]].




\bibitem{Linde:2007fr}
  A.~Linde,
  ``Inflationary Cosmology,''
  arXiv:0705.0164 [hep-th].


\bibitem{Freese:1990rb}
  K.~Freese, J.~A.~Frieman and A.~V.~Olinto,
 ``Natural inflation with pseudo - Nambu-Goldstone bosons,''
  Phys.\ Rev.\ Lett.\  {\bf 65}, 3233 (1990);
  F.~C.~Adams, J.~R.~Bond, K.~Freese, J.~A.~Frieman and A.~V.~Olinto,
 ``Natural Inflation: Particle Physics Models, Power Law Spectra For Large
 Scale Structure, And Constraints From Cobe,''
  Phys.\ Rev.\  D {\bf 47}, 426 (1993)
  [arXiv:hep-ph/9207245].



\bibitem{Savage:2006tr}
  C.~Savage, K.~Freese and W.~H.~Kinney,
 ``Natural inflation: Status after WMAP 3-year data,''
  Phys.\ Rev.\  D {\bf 74}, 123511 (2006)
  [arXiv:hep-ph/0609144].






\bibitem{Lyth:1996im}
  D.~H.~Lyth,
``What would we learn by detecting a gravitational wave signal in the  cosmic
microwave background anisotropy?,''
  Phys.\ Rev.\ Lett.\  {\bf 78}, 1861 (1997)
  [arXiv:hep-ph/9606387];
  D.~H.~Lyth,
``Particle physics models of inflation,''
  arXiv:hep-th/0702128.


\bibitem{Liddle:1998jc}
  A.~R.~Liddle, A.~Mazumdar and F.~E.~Schunck,
  ``Assisted inflation,''
  Phys.\ Rev.\  D {\bf 58}, 061301 (1998)
  [arXiv:astro-ph/9804177];
  P.~Kanti and K.~A.~Olive,
  ``Assisted chaotic inflation in higher dimensional theories,''
  Phys.\ Lett.\  B {\bf 464}, 192 (1999)
  [arXiv:hep-ph/9906331].


  \bibitem{Kachru:unpub}
S. Kachru, talk at 2005 Solvay conference; talk at COSMO 2006 conference
http://cosmo06.ucdavis.edu/talks/Kachru.pdf

\bibitem{Kallosh:2007ig}
  R.~Kallosh,
  ``On Inflation in String Theory,''
  arXiv:hep-th/0702059.


\bibitem{Baumann:2006cd}
  D.~Baumann and L.~McAllister,
``A microscopic limit on gravitational waves from D-brane inflation,''
  arXiv:hep-th/0610285.



\bibitem{Bean:2007hc}
  R.~Bean, S.~E.~Shandera, S.~H.~Henry Tye and J.~Xu,
``Comparing brane inflation to WMAP,''
  arXiv:hep-th/0702107.


  \bibitem{Hybrid}
A.~D.~Linde, ``Axions in inflationary cosmology,'' Phys.\ Lett.\ B
{\bf 259}, 38 (1991); A.~D.~Linde, ``Hybrid inflation,'' Phys.\
Rev.\ D {\bf 49}, 748 (1994) [astro-ph/9307002].

\bibitem{Kachru:2003sx}
  S.~Kachru, R.~Kallosh, A.~Linde, J.~M.~Maldacena, L.~McAllister and S.~P.~Trivedi,
  ``Towards inflation in string theory,''
  JCAP {\bf 0310}, 013 (2003)
  [arXiv:hep-th/0308055].

\bibitem{Kachru:2003aw}
  S.~Kachru, R.~Kallosh, A.~Linde and S.~P.~Trivedi,
  ``De Sitter vacua in string theory,''
  Phys.\ Rev.\  D {\bf 68}, 046005 (2003)
  [arXiv:hep-th/0301240].


\bibitem{Kallosh:2004yh}
  R.~Kallosh and A.~Linde,
  ``Landscape, the scale of SUSY breaking, and inflation,''
  JHEP {\bf 0412} (2004) 004
  [arXiv:hep-th/0411011].




\bibitem{Binetruy:1986ss}
  P.~Binetruy and M.~K.~Gaillard,
  ``Candidates For The Inflaton Field In Superstring Models,''
  Phys.\ Rev.\  D {\bf 34} (1986) 3069.

\bibitem{Gaillard:1995az}
  M.~K.~Gaillard, H.~Murayama and K.~A.~Olive,
  ``Preserving flat directions during inflation,''
  Phys.\ Lett.\  B {\bf 355}, 71 (1995)
  [arXiv:hep-ph/9504307].


\bibitem{Banks:2003sx}
  T.~Banks, M.~Dine, P.~J.~Fox and E.~Gorbatov,
  ``On the possibility of large axion decay constants,''
  JCAP {\bf 0306}, 001 (2003)
  [arXiv:hep-th/0303252].


\bibitem{Kim:2004rp}
  J.~E.~Kim, H.~P.~Nilles and M.~Peloso,
  ``Completing natural inflation,''
  JCAP {\bf 0501}, 005 (2005)
  [arXiv:hep-ph/0409138].




\bibitem{Dimopoulos:2005ac}
  S.~Dimopoulos, S.~Kachru, J.~McGreevy and J.~G.~Wacker,
  ``N-flation,''
  arXiv:hep-th/0507205;

\bibitem{Easther:2005zr}
  R.~Easther and L.~McAllister,
  ``Random matrices and the spectrum of N-flation,''
  JCAP {\bf 0605}, 018 (2006)
  [arXiv:hep-th/0512102];
  M.~E.~Olsson,
  ``Inflation Assisted by Heterotic Axions,''
  JCAP {\bf 0704}, 019 (2007)
  [arXiv:hep-th/0702109].

\bibitem{Green:2007gs}
  D.~Green,
  ``Reheating in UV Complete Theories,''
  arXiv:0707.3832 [hep-th].

\bibitem{ArkaniHamed:2003wu}
  N.~Arkani-Hamed, H.~C.~Cheng, P.~Creminelli and L.~Randall,
 ``Extranatural inflation,''
      Phys.\ Rev.\ Lett.\  {\bf 90}, 221302 (2003)
        [arXiv:hep-th/0301218];
  N.~Arkani-Hamed, H.~C.~Cheng, P.~Creminelli and L.~Randall,
 ``Pseudonatural inflation,''
      JCAP {\bf 0307}, 003 (2003)
        [arXiv:hep-th/0302034].




\bibitem{Kawasaki:2000yn}
  M.~Kawasaki, M.~Yamaguchi and T.~Yanagida,
  ``Natural chaotic inflation in supergravity,''
  Phys.\ Rev.\ Lett.\  {\bf 85}, 3572 (2000)
  [arXiv:hep-ph/0004243].






\bibitem{Sasaki:1995aw}
  M.~Sasaki and E.~D.~Stewart,
  ``A General Analytic Formula For The Spectral Index Of The Density
  Perturbations Produced During Inflation,''
  Prog.\ Theor.\ Phys.\  {\bf 95}, 71 (1996)
  [arXiv:astro-ph/9507001].








\bibitem{Svrcek:2006yi}
  P.~Svrcek and E.~Witten,
  ``Axions in string theory,''
  JHEP {\bf 0606}, 051 (2006)
  [arXiv:hep-th/0605206].

\bibitem{Becker:2007zj}
  K.~Becker, M.~Becker and J.~H.~Schwarz,
  ``String theory and M-theory: A modern introduction,''
{\it  Cambridge, UK: Cambridge Univ. Pr. (2007) 739 p}


\bibitem{Polchinski:1998rr}
  J.~Polchinski,
  ``String theory. Vol. 2: Superstring theory and beyond,''
{\it  Cambridge, UK: Univ. Pr. (1998) 531 p}

\bibitem{Denef:2004dm}
  F.~Denef, M.~R.~Douglas and B.~Florea,
  ``Building a better racetrack,''
  JHEP {\bf 0406}, 034 (2004)
  [arXiv:hep-th/0404257].

\bibitem{Becker:2002nn}
  K.~Becker, M.~Becker, M.~Haack and J.~Louis,
  ``Supersymmetry breaking and alpha'-corrections to flux induced
  potentials,''
  JHEP {\bf 0206}, 060 (2002)
  [arXiv:hep-th/0204254].





\bibitem{Blanco-Pillado:2005fn}
  J.~J.~Blanco-Pillado, R.~Kallosh and A.~Linde,
  ``Supersymmetry and stability of flux vacua,''
  JHEP {\bf 0605}, 053 (2006)
  [arXiv:hep-th/0511042];
  A.~Ceresole, G.~Dall'Agata, A.~Giryavets, R.~Kallosh and A.~Linde,
  ``Domain walls, near-BPS bubbles, and probabilities in the landscape,''
  Phys.\ Rev.\  D {\bf 74}, 086010 (2006)
  [arXiv:hep-th/0605266].



\bibitem{Blanco-Pillado:2006he}
  J.~J.~Blanco-Pillado, C.~P. ~Burgess, J.~M. ~Cline, C. ~Escoda, M. ~Gomez-Reino, R. ~Kallosh, A. ~Linde, F. ~Quevedo,
  ``Inflating in a better racetrack,''
  JHEP {\bf 0609}, 002 (2006)
  [arXiv:hep-th/0603129].


\bibitem{Candelas:1990pi}
  P.~Candelas and X.~de la Ossa, ``Moduli space of Calabi-Yau manifolds,''
  Nucl.\ Phys.\  B {\bf 355}, 455 (1991).

\bibitem{Candelas:1990rm}
  P.~Candelas, X.~C.~De La Ossa, P.~S.~Green and L.~Parkes,
  ``A pair of Calabi-Yau manifolds as an exactly soluble superconformal
  theory,''
  Nucl.\ Phys.\  B {\bf 359}, 21 (1991).

\bibitem{Hosono:1994ax}
  S.~Hosono, A.~Klemm, S.~Theisen and S.~T.~Yau,
 ``Mirror symmetry, mirror map and applications to complete intersection
  Calabi-Yau spaces,''
  Nucl.\ Phys.\  B {\bf 433}, 501 (1995)
  [arXiv:hep-th/9406055].

\bibitem{Hosono:1994av}
  S.~Hosono, A.~Klemm and S.~Theisen,
  ``Lectures On Mirror Symmetry,''
  arXiv:hep-th/9403096.





\bibitem{Berg:2007wt}
  M.~Berg, M.~Haack and E.~Pajer,
  ``Jumping Through Loops: On Soft Terms from Large Volume Compactifications,''
  arXiv:0704.0737 [hep-th];
  M.~Berg, M.~Haack and B.~Kors,
  ``On volume stabilization by quantum corrections,''
  Phys.\ Rev.\ Lett.\  {\bf 96}, 021601 (2006)
  [arXiv:hep-th/0508171].
  G.~von Gersdorff and A.~Hebecker,
  Kaehler corrections for the volume modulus of flux compactifications,''
  Phys.\ Lett.\  B {\bf 624}, 270 (2005)
  [arXiv:hep-th/0507131].





\bibitem{Balasubramanian:2004uy}
  V.~Balasubramanian and P.~Berglund,
  ``Stringy corrections to Kaehler potentials, SUSY breaking, and the
  cosmological constant problem,''
  JHEP {\bf 0411}, 085 (2004)
  [arXiv:hep-th/0408054];
  V.~Balasubramanian, P.~Berglund, J.~P.~Conlon and F.~Quevedo,
 ``Systematics of moduli stabilisation in Calabi-Yau flux
  compactifications,''
  JHEP {\bf 0503}, 007 (2005)
  [arXiv:hep-th/0502058].

\bibitem{Bobkov:2004cy}
  K.~Bobkov,
  ``Volume stabilization via alpha' corrections in type IIB theory with
  fluxes,''
  JHEP {\bf 0505}, 010 (2005)
  [arXiv:hep-th/0412239].

\bibitem{Cicoli:2007xp}
  M.~Cicoli, J.~P.~Conlon and F.~Quevedo,
  ``Systematics of String Loop Corrections in Type IIB Calabi-Yau Flux
  Compactifications,''
  arXiv:0708.1873 [hep-th].

\bibitem{Douglas:2006es}
  M.~R.~Douglas and S.~Kachru,
  ``Flux compactification,''
  arXiv:hep-th/0610102.


\bibitem{deWit:1984pk}
  B.~de Wit and A.~Van Proeyen,
  ``Potentials And Symmetries Of General Gauged N=2 Supergravity: Yang-Mills
 Models,''
  Nucl.\ Phys.\  B {\bf 245}, 89 (1984).

\bibitem{Gunaydin:1983bi}
  M.~Gunaydin, G.~Sierra and P.~K.~Townsend,
  ``The Geometry Of N=2 Maxwell-Einstein Supergravity And Jordan Algebras,''
  Nucl.\ Phys.\  B {\bf 242}, 244 (1984).

\bibitem{Denef:2004ze}
  F.~Denef and M.~R.~Douglas,
  ``Distributions of flux vacua,''
  JHEP {\bf 0405}, 072 (2004)
  [arXiv:hep-th/0404116];
  F.~Denef and M.~R.~Douglas,
  ``Distributions of nonsupersymmetric flux vacua,''
  JHEP {\bf 0503}, 061 (2005)
  [arXiv:hep-th/0411183].

\bibitem{Soroush:2007ed}
  M.~Soroush,
  ``Constraints on meta-stable de Sitter flux vacua,''
  [arXiv:hep-th/0702204].

\bibitem{Gomez-Reino:2006dk}
  M.~Gomez-Reino and C.~A.~Scrucca,
  ``Locally stable non-supersymmetric Minkowski vacua in supergravity,''
  JHEP {\bf 0605}, 015 (2006)
  [arXiv:hep-th/0602246];
  M.~Gomez-Reino and C.~A.~Scrucca,
  ``Constraints for the existence of flat and stable non-supersymmetric vacua
  in supergravity,''
  JHEP {\bf 0609}, 008 (2006)
  [arXiv:hep-th/0606273].







\bibitem{Conlon:2005ki}
  J.~P.~Conlon, F.~Quevedo and K.~Suruliz,
  ``Large-volume flux compactifications: Moduli spectrum and D3/D7 soft
  supersymmetry breaking,''
  JHEP {\bf 0508}, 007 (2005)
  [arXiv:hep-th/0505076].

\bibitem{Westphal:2006tn}
  A.~Westphal,
``de Sitter string vacua from Kaehler uplifting,''
  JHEP {\bf 0703}, 102 (2007)
  [arXiv:hep-th/0611332].

\bibitem{Aspinwall:2005ad}
  P.~S.~Aspinwall and R.~Kallosh,
  ``Fixing all moduli for M-theory on K3 x K3,''
  JHEP {\bf 0510}, 001 (2005)
  [arXiv:hep-th/0506014].


\bibitem{Tripathy:2002qw}
  P.~K.~Tripathy and S.~P.~Trivedi,
  ``Compactification with flux on K3 and tori,''
  JHEP {\bf 0303}, 028 (2003)
  [arXiv:hep-th/0301139].



\bibitem{Andrianopoli:2003jf}
  L.~Andrianopoli, R.~D'Auria, S.~Ferrara and M.~A.~Lledo,
  ``4-D gauged supergravity analysis of type IIB vacua on K3 x T**2/Z(2),''
  JHEP {\bf 0303}, 044 (2003)
  [arXiv:hep-th/0302174].




\bibitem{Bergshoeff:2005yp}
  E.~Bergshoeff, R.~Kallosh, A.~K.~Kashani-Poor, D.~Sorokin and A.~Tomasiello,
 ``An index for the Dirac operator on D3 branes with background fluxes,''
  JHEP {\bf 0510}, 102 (2005)
  [arXiv:hep-th/0507069];
  R.~Kallosh, A.~K.~Kashani-Poor and A.~Tomasiello,
  ``Counting fermionic zero modes on M5 with fluxes,''
  JHEP {\bf 0506}, 069 (2005)
  [arXiv:hep-th/0503138];
  N.~Saulina,
  ``Topological constraints on stabilized flux vacua,''
  Nucl.\ Phys.\  B {\bf 720}, 203 (2005)
  [arXiv:hep-th/0503125].



\bibitem{Acharya:2007rc}
  B.~S.~Acharya, K.~Bobkov, G.~L.~Kane, P.~Kumar and J.~Shao,
  ``Explaining the electroweak scale and stabilizing moduli in M theory,''
  arXiv:hep-th/0701034.

\bibitem{Grimm} Thomas Grimm, private communication;
  T.~W.~Grimm,
  arXiv:0710.3883 [hep-th].

\bibitem{Grimm:2004uq}
  T.~W.~Grimm and J.~Louis,
  ``The effective action of N = 1 Calabi-Yau orientifolds,''
  Nucl.\ Phys.\  B {\bf 699}, 387 (2004)
  [arXiv:hep-th/0403067].
  H.~Jockers and J.~Louis,
  ``The effective action of D7-branes in N = 1 Calabi-Yau orientifolds,''
  Nucl.\ Phys.\  B {\bf 705}, 167 (2005)
  [arXiv:hep-th/0409098].
  H.~Jockers and J.~Louis,
  ``D-terms and F-terms from D7-brane fluxes,''
  Nucl.\ Phys.\  B {\bf 718}, 203 (2005)
  [arXiv:hep-th/0502059].
  T.~W.~Grimm,
  ``Non-Perturbative Corrections and Modularity in N=1 Type IIB
  Compactifications,''
  arXiv:0705.3253 [hep-th].


\bibitem{Grana:2005ny}
  M.~Grana, J.~Louis and D.~Waldram,
  ``Hitchin functionals in N = 2 supergravity,''
  JHEP {\bf 0601}, 008 (2006)
  [arXiv:hep-th/0505264];
  M.~Grana, R.~Minasian, M.~Petrini and A.~Tomasiello,
  ``A scan for new N=1 vacua on twisted tori,''
  JHEP {\bf 0705}, 031 (2007)
  [arXiv:hep-th/0609124];
  M.~Grana, J.~Louis and D.~Waldram,
  ``SU(3) x SU(3) compactification and mirror duals of magnetic fluxes,''
  JHEP {\bf 0704}, 101 (2007)
  [arXiv:hep-th/0612237].

\bibitem{Giddings:2001yu}
  S.~B.~Giddings, S.~Kachru and J.~Polchinski,
  ``Hierarchies from fluxes in string compactifications,''
      Phys.\ Rev.\  D {\bf 66}, 106006 (2002)
        [arXiv:hep-th/0105097].





\bibitem{BlancoPillado:2004ns}
  J.~J.~Blanco-Pillado, C.~P.~ Burgess, J.~M.~ Cline ,C.~ Escoda, M.~ Gomez-Reino, R.~ Kallosh, A.~ Linde, F.~ Quevedo,
 ``Racetrack inflation,''
  JHEP {\bf 0411}, 063 (2004)
  [arXiv:hep-th/0406230].

\bibitem{Alishahiha:2004eh}
  M.~Alishahiha, E.~Silverstein and D.~Tong,
 ``DBI in the sky,''
  Phys.\ Rev.\  D {\bf 70}, 123505 (2004)
  [arXiv:hep-th/0404084].


\bibitem{Krause:2007jr}
  A.~Krause,
  ``Large Gravitational Waves and Lyth Bound in Multi Brane Inflation,''
  arXiv:0708.4414 [hep-th].






\bibitem{Becker:2007ui}
  M.~Becker, L.~Leblond and S.~E.~Shandera,
 ``Inflation from Wrapped Branes,''
  arXiv:0709.1170 [hep-th].

\bibitem{Kobayashi:2007hm}
  T.~Kobayashi, S.~Mukohyama and S.~Kinoshita,
  ``Constraints on Wrapped DBI Inflation in a Warped Throat,''
  arXiv:0708.4285 [hep-th].



\end{thebibliography}
\end{document}